\documentclass[prd,aps,nofootinbib,floatfix,10pt]{revtex4}
\usepackage{amsmath,graphicx,epsfig,amssymb}
\usepackage{inputenc}
\usepackage[usenames]{color}
\usepackage{ulem} 
\usepackage{bigstrut}
\usepackage{slashed}
\usepackage{multirow}
\usepackage{subfigure}

\allowdisplaybreaks

\begin{document}
	
\title{QCD sum rules analysis of weak decays of doubly heavy baryons: the $b\to c$ processes}
	
\author{Zhi-Peng Xing$^{2}$~~\footnote{Email:zpxing@sjtu.edu.cn} Zhen-Xing Zhao$^{1}$~\footnote{Email:zhaozx19@imu.edu.cn} }
\affiliation{$^{1}$ School of Physical Science and Technology,
        Inner Mongolia University, Hohhot 010021, China }
\affiliation{$^{2}$ Tsung-Dao Lee Institute, Shanghai Jiao-Tong University, Shanghai
		200240, China }

\begin{abstract}
A comprehensive study of $b\to c$ weak decays
of doubly heavy baryons is presented in this paper. The transition form factors
as well as the pole residues of the initial and final states are respectively
obtained by investigating the three-point and two-point correlation
functions in QCD sum rules. Contributions from up to dimension-6 operators
are respectively considered for the two-point and three-point correlation
functions. The obtained form factors are then applied to a phenomenological
analysis of semi-leptonic decays.
\end{abstract}
\maketitle
	
\section{Introduction}

Quark model has achieved brilliant success in the study of hadron
spectroscopy. However, the existence of doubly heavy baryons had become
a long-standing problem in experiments until LHCb reported the observation
of $\Xi_{cc}^{++}$ \cite{LHCb:2017iph}. The discovery of the doubly charmed baryon has triggered many related theoretical researches on the masses, lifetimes, strong coupling constants, and decay widths of doubly heavy baryons. They are based on various model calculations\cite{Yu:2017zst,Luchinsky:2020fdf,Gerasimov:2019jwp,Wang:2017mqp,Meng:2017udf,Gutsche:2017hux,Xiao:2017udy,Lu:2017meb,Xiao:2017dly,Zhao:2018mrg,Xing:2018lre,Dhir:2018twm,Jiang:2018oak,Gutsche:2018msz,Gutsche:2019wgu,Yu:2019yfr,Gutsche:2019iac,Berezhnoy:2019zwu,Ke:2019lcf,Cheng:2020wmk,Hu:2020mxk,Rahmani:2020pol,Ivanov:2020xmw,Li:2020qrh,Han:2021gkl}, SU(3) symmetry analysis\cite{Wang:2017azm,Shi:2017dto,Zhang:2018llc}, effective theories\cite{Berezhnoy:2018bde,Li:2017pxa,Guo:2017vcf,Ma:2017nik,Yao:2018zze,Yao:2018ifh,Meng:2018zbl,Shi:2020qde,Qiu:2020omj,Olamaei:2021hjd,Qin:2021dqo},  QCD sum rules\cite{Hu:2017dzi,Li:2018epz,Shi:2019hbf}and light-cone sum rules \cite{Cui:2017udv,Shi:2019fph,Hu:2019bqj,Olamaei:2020bvw,Alrebdi:2020rev,Rostami:2020euc,Aliev:2020aon,Aliev:2021hqq,Azizi:2020zin}. For a recent review, see~\cite{Yu:2019lxw}. It is promising that more doubly heavy baryons will be discovered in the near future.

In~\cite{Wang:2017mqp}, we performed an analysis of weak
decays of doubly heavy baryons using the approach of light-front quark
model. However, model-dependent parameters are inevitably introduced.
In view of this, in \cite{Shi:2019hbf}, we investigated the weak decays of doubly heavy baryons to singly heavy baryons
using QCD sum rules (QCDSR). QCDSR is a QCD-based approach to deal
with the hadron parameters. It reveals a connection between hadron
phenomenology and QCD vacuum via a few universal condensate parameters.
However, the processes induced by the $b\to c$ transition were not
considered in \cite{Shi:2019hbf}. In particular, these processes
are considered to be important for the search of other doubly heavy
baryons. This work aims to fill this gap. Specifically, we will consider
the following processes ($q=u/d$):
\begin{itemize}
\item the $bb$ sector,
\begin{align*}
\begin{array}{lcl}
\Xi_{bb}(bbq) & \to & \Xi_{bc}(bcq),\\
\Omega_{bb}(bbs) & \to & \Omega_{bc}(bcs),
\end{array}
\end{align*}
\item the $bc$ sector,
\begin{align*}
\begin{array}{lcl}
\Xi_{bc}(bcq) & \to & \Xi_{cc}(ccq),\\
\Omega_{bc}(bcs) & \to & \Omega_{cc}(ccs).
\end{array}
\end{align*}
\end{itemize}

The transition matrix element can be parametrized by the so-called
helicity form factors $f_{0,+,\perp}$ and $g_{0,+,\perp}$ \cite{Feldmann:2011xf}:
\begin{eqnarray}
\langle\mathcal{B}_{2}(P_{2})|(V-A)_{\mu}|\mathcal{B}_{1}(P_{1})\rangle & = & \bar{u}(P_{2},s_{2})\bigg[\frac{q_{\mu}}{q^{2}}(M_{1}-M_{2})f_{0}(q^{2})+\frac{M_{1}+M_{2}}{Q_{+}}((P_{1}+P_{2})_{\mu}-(M_{1}^{2}-M_{2}^{2})\frac{q_{\mu}}{q^{2}})f_{+}(q^{2})\nonumber \\
 &  & +(\gamma_{\mu}-\frac{2M_{2}}{Q_{+}}P_{1\mu}-\frac{2M_{1}}{Q_{+}}P_{2\mu})f_{\perp}(q^{2})\bigg]u(P_{1},s_{1})\nonumber \\
 & - & \bar{u}(P_{2},s_{2})\gamma_{5}\bigg[\frac{q_{\mu}}{q^{2}}(M_{1}+M_{2})g_{0}(q^{2})+\frac{M_{1}-M_{2}}{Q_{-}}((P_{1}+P_{2})_{\mu}-(M_{1}^{2}-M_{2}^{2})\frac{q_{\mu}}{q^{2}})g_{+}(q^{2})\nonumber \\
 &  & +(\gamma_{\mu}+\frac{2M_{2}}{Q_{-}}P_{1\mu}-\frac{2M_{1}}{Q_{-}}P_{2\mu})g_{\perp}(q^{2})\bigg]u(P_{1},s_{1})
\end{eqnarray}
with $Q_{\pm}=(M_{1}\pm M_{2})^{2}-q^{2}$. These form factors can be extracted using the three-point correlation
functions in QCDSR.

The leading logarithmic corrections are also considered in this work.
In some literatures, the anomalous dimensions of interpolating currents
of baryons are incorrectly cited. Therefore, in \cite{Zhao:2021xwl},
we calculated these anomalous dimensions at one-loop level.

It is worth noting that Heavy Quark Effective Theory (HQET) does not
apply to the situation of doubly heavy baryons. However, the heavy
quark limit can still be taken from the full theory results, as can be seen in
\cite{Shuryak:1981fza,MarquesdeCarvalho:1999bqs,Zhao:2020wbw}. Some
efforts were made to develop the effective theory for doubly heavy
baryons in \cite{Shi:2020qde}.

The rest of this paper is arranged as follows. In Sec. II, the QCDSR
methods for the two-point and three-point correlation functions are
briefly introduced, and corresponding numerical results are shown
in Sec. III. The obtained form factors are applied to phenomenology
analysis in Sec. IV. A short summary is given in the last section.

\section{QCD sum rules}

\subsection{The two-point correlation functions}

The pole residue of the doubly heavy baryon $\mathcal{B}_{Q_{1}Q_{2}q_{3}}$ can be obtained by calculating the following two-point
correlation function
\begin{equation}
\Pi(q)=i\int d^{4}xe^{iq\cdot x}\langle0|T\left[J_{\mathcal{B}_{Q_{1}Q_{2}q_{3}}}(x)\bar{J}_{\mathcal{B}_{Q_{1}Q_{2}q_{3}}}(0)\right]|0\rangle.\label{eq:two-point}
\end{equation}
The interpolating currents of doubly heavy baryons are
\begin{equation}
J_{\mathcal{B}_{QQq}}(y)=\epsilon_{abc}(Q_{a}^{T}C\gamma^{\mu}Q_{b})\gamma_{\mu}\gamma_{5}q_{c},\quad J_{\mathcal{B}_{Q_{1}Q_{2}q}}(y)=\epsilon_{abc}\frac{1}{\sqrt{2}}(Q_{1a}^{T}C\gamma^{\mu}Q_{2b}+Q_{2a}^{T}C\gamma^{\mu}Q_{1b})\gamma_{\mu}\gamma_{5}q_{c}.
\end{equation}
At the hadron level, by inserting the complete set of baryons in Eq.
(\ref{eq:two-point}), one can obtain
\begin{equation}
\Pi^{{\rm had}}(q)=\lambda_{+}^{2}\frac{\not\!q+M_{+}}{M_{+}^{2}-q^{2}}+\lambda_{-}^{2}\frac{\not\!q-M_{-}}{M_{-}^{2}-q^{2}}+\cdots,
\end{equation}
where we have also considered the contribution from the negative-parity
baryon, and $M_{\pm}$ ($\lambda_{\pm}$) are respectively the masses
(pole residues) of positive- and negative-parity baryons. The
pole residues are introduced as
\begin{align}
\langle0|J_{+}(0)|\mathcal{B}_{+}(p,s)\rangle & =u(p,s)\lambda_{+},\nonumber \\
\langle0|J_{+}(0)|\mathcal{B}_{-}(p,s)\rangle & =(i\gamma_{5})u(p,s)\lambda_{-}.
\end{align}

At the QCD level, the correlation functions are calculated using the
operator product expansion (OPE) technique.
Contributions from up to dimension-5 operators are considered in this work.
The result can be formally
written as
\begin{equation}
\Pi(q)=\not\!q\Pi_{1}(q^{2})+\Pi_{2}(q^{2}).
\end{equation}
$\Pi_{i}$ can be written in terms of dispersion relation for practical
purpose
\begin{equation}
\Pi_{i}(q^{2})=\int_{0}^{\infty}ds\frac{\rho_{i}(s)}{s-q^{2}}.
\end{equation}
Assuming quark-hadron duality and performing the Borel transformation,
one can obtain the following sum rule for $1/2^{+}$ baryon
\begin{equation}
(M_{+}+M_{-})\lambda_{+}^{2}e^{-M_{+}^{2}/T_{+}^{2}}=\int^{s_{+}}ds(M_{-}\rho_{1}(s)+\rho_{2}(s))e^{-s/T_{+}^{2}},\label{eq:two-point_sum_rule}
\end{equation}
where $T_{+}^{2}$ and $s_{+}$ are respectively the Borel parameter
and continuum threshold parameter. From Eq. (\ref{eq:two-point_sum_rule}),
one can obtain the squared mass for $1/2^{+}$ baryon
\begin{equation}
M_{+}^{2}=\frac{\int^{s_{+}}ds\ (M_{-}\rho_{1}+\rho_{2})\ s\ e^{-s/T_{+}^{2}}}{\int^{s_{+}}ds\ (M_{-}\rho_{1}+\rho_{2})\ e^{-s/T_{+}^{2}}}.
\end{equation}

The leading logarithmic (LL) corrections are considered in
this work. The Wilson coefficients of OPE should be multiplied by
\begin{equation}
\Bigg(\frac{{\rm log}(\mu_{0}/\Lambda_{{\rm QCD}}^{(n_{f})})}{{\rm log}(\mu/\Lambda_{{\rm QCD}}^{(n_{f})})}\Bigg)^{2\gamma_{J}-\gamma_{O}},
\end{equation}
where $\gamma_{J}$ and $\gamma_{O}$ are anomalous dimensions of
the interpolating current and the local operator respectively. $\Lambda_{{\rm QCD}}^{(n_{f})}$
is given by $\Lambda_{{\rm QCD}}^{(3)}=223\ {\rm MeV}$ and $\Lambda_{{\rm QCD}}^{(4)}=170\ {\rm MeV}$
\cite{Buras:1998raa,Zhao:2020mod}. The renormalization scale $\mu_{0}\sim1\ {\rm GeV}$, and
$\mu$ is chosen as $m_{c}$ for doubly charmed baryons and $m_{b}$
for doubly bottom and bottom-charmed baryons. The masses and pole residues of doubly heavy baryons are also be considered in~\cite{Wang:2018lhz}.

\subsection{The three-point correlation functions}

The following three-point correlation functions are adopted to extract
the transition form factors of ${\cal B}_{bQq}\to{\cal B}_{cQq}$
\begin{equation}
\Pi_{\mu}^{V,A}(P_{1},P_{2})=i^{2}\int d^{4}xd^{4}ye^{-iP_{1}\cdot x+iP_{2}\cdot y}\langle0|T\{J_{\mathcal{B}_{cQq}}(y)(V_{\mu},A_{\mu})(0)\bar{J}_{\mathcal{B}_{bQq}}(x)\}|0\rangle.\label{eq:three-point}
\end{equation}
At the hadron level, the complete sets of baryon states are inserted
to the correlation function to obtain for the vector current correlation function
\begin{eqnarray}
\Pi_{\mu}^{V,{\rm had}}(P_{1},P_{2}) & = & \frac{\lambda_{f}^{+}\lambda_{i}^{+}}{(P_{2}^{2}-M_{2}^{+2})(P_{1}^{2}-M_{1}^{+2})}(\not\!P_{2}+M_{2}^{+}){\cal V}_{\mu}^{++}(\not\!P_{1}+M_{1}^{+})\nonumber \\
 & + & \frac{\lambda_{f}^{+}\lambda_{i}^{-}}{(P_{2}^{2}-M_{2}^{+2})(P_{1}^{2}-M_{1}^{-2})}(\not\!P_{2}+M_{2}^{+}){\cal V}_{\mu}^{+-}(\not\!P_{1}-M_{1}^{-})\nonumber \\
 & + & \frac{\lambda_{f}^{-}\lambda_{i}^{+}}{(P_{2}^{2}-M_{2}^{-2})(P_{1}^{2}-M_{1}^{+2})}(\not\!P_{2}-M_{2}^{-}){\cal V}_{\mu}^{-+}(\not\!P_{1}+M_{1}^{+})\nonumber \\
 & + & \frac{\lambda_{f}^{-}\lambda_{i}^{-}}{(P_{2}^{2}-M_{2}^{-2})(P_{1}^{2}-M_{1}^{-2})}(\not\!P_{2}-M_{2}^{-}){\cal V}_{\mu}^{--}(\not\!P_{1}-M_{1}^{-})\nonumber \\
 & + & ...\label{eq:hadron}
\end{eqnarray}
where
\begin{eqnarray}
{\cal V}_{\mu}^{ij} & \equiv & \frac{q_{\mu}}{q^{2}}(M_{1}^{j}-M_{2}^{i})f_{0}^{ij}(q^{2})+\frac{M_{1}^{j}+M_{2}^{i}}{Q_{+}}((P_{1}+P_{2})_{\mu}-(M_{1}^{j2}-M_{2}^{i2})\frac{q_{\mu}}{q^{2}})f_{+}^{ij}(q^{2})\nonumber \\
 &  & +(\gamma_{\mu}-\frac{2M_{2}^{j}}{Q_{+}}P_{1\mu}-\frac{2M_{1}^{i}}{Q_{+}}P_{2\mu})f_{\perp}^{ij}(q^{2})
\end{eqnarray}
with $i,j=+,-$. In this step, both of the contributions from positive-
and negative-parity baryons are considered. $M_{1(2)}^{+(-)}$ and
$\lambda_{i(f)}^{+(-)}$ respectively denote the mass and pole residue
of the baryon in the initial (final) state with positive (negative)
parity, and $f_{0,+,\perp}^{ij}(q^{2})$ are 12 form factors defined
by:
\begin{align}
\langle{\cal B}_{f}^{+}(p_{2},s_{2})|V_{\mu}|{\cal B}_{i}^{+}(p_{1},s_{1})\rangle & =\bar{u}_{{\cal B}_{f}^{+}}(p_{2},s_{2}){\cal V}_{\mu}^{++}u_{{\cal B}_{i}^{+}}(p_{1},s_{1}),\nonumber \\
\langle{\cal B}_{f}^{+}(p_{2},s_{2})|V_{\mu}|{\cal B}_{i}^{-}(p_{1},s_{1})\rangle & =\bar{u}_{{\cal B}_{f}^{+}}(p_{2},s_{2}){\cal V}_{\mu}^{+-}(i\gamma_{5})u_{{\cal B}_{i}^{-}}(p_{1},s_{1}),\nonumber \\
\langle{\cal B}_{f}^{-}(p_{2},s_{2})|V_{\mu}|{\cal B}_{i}^{+}(p_{1},s_{1})\rangle & =\bar{u}_{{\cal B}_{f}^{-}}(p_{2},s_{2})(i\gamma_{5}){\cal V}_{\mu}^{-+}u_{{\cal B}_{i}^{+}}(p_{1},s_{1}),\nonumber \\
\langle{\cal B}_{f}^{-}(p_{2},s_{2})|V_{\mu}|{\cal B}_{i}^{-}(p_{1},s_{1})\rangle & =\bar{u}_{{\cal B}_{f}^{-}}(p_{2},s_{2})(i\gamma_{5}){\cal V}_{\mu}^{--}(i\gamma_{5})u_{{\cal B}_{i}^{-}}(p_{1},s_{1}).
\end{align}

\begin{figure}[htbp!]
\includegraphics[width=0.6\columnwidth]{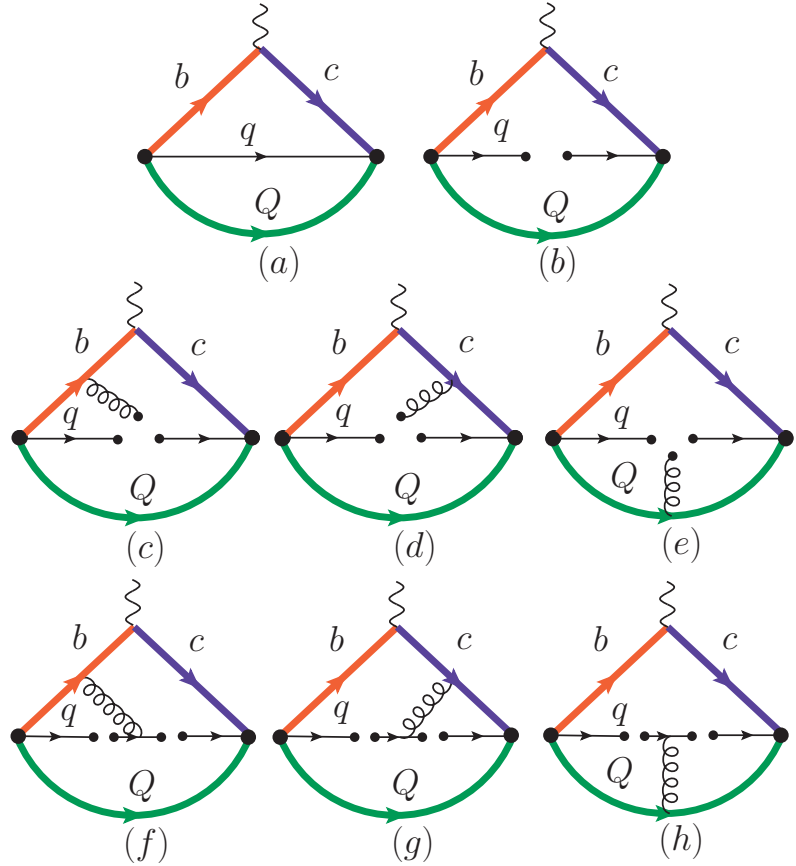} \caption{Feynman diagrams in the calculation of the three-point correlation
functions at the quark level.}
\label{qcdsr}
\end{figure}

At the quark level, the correlation functions in Eq. (\ref{eq:three-point}) are calculated using
OPE technique. In this work, contributions from the perturbative term
(dim-0), quark condensate term (dim-3), mixed quark-gluon condensate
term (dim-5), and four-quark condensate term (dim-6) are considered,
as can be seen in Fig.~\ref{qcdsr}. The vector current correlation function is
further written into the double dispersion relation
\begin{equation}
\Pi_{\mu}^{V}(P_{1},P_{2})=\int^{\infty}ds_{1}\int^{\infty}ds_{2}\frac{\rho_{\mu}^{V}(s_{1},s_{2},q^{2})}{(s_{1}-P_{1}^{2})(s_{2}-P_{2}^{2})},\quad q=P_{1}-P_{2},\label{eq:quark}
\end{equation}
where the spectral density functions $\rho_{\mu}^{V}(s_{1},s_{2},q^{2})$
are obtained by taking discontinuities for $s_{1}$ and $s_{2}$.
Our method is further illustrated by the calculation of perturbative
diagram below.

The spectral density of the perturbative diagram in Fig.~\ref{qcdsr}(a)
can be obtained as
\begin{eqnarray}
\rho_{\mu}^{V,{\rm pert}}(P_{1}^{2},P_{2}^{2},q^{2}) & = & \frac{1}{(2\pi i)^{2}}\int\frac{d^{4}p_{b}d^{4}p_{c}d^{4}p_{Q}d^{4}p_{q}}{(2\pi)^{16}}N_{\mu}\nonumber \\
 & \times & (2\pi)^{8}\delta^{4}(p_{b}+p_{Q}+p_{q}-P_{1})\delta^{4}(p_{c}+p_{Q}+p_{q}-P_{2})\nonumber \\
 & \times & (-2\pi i)^{4}\delta(p_{b}^{2}-m_{b}^{2})\delta(p_{c}^{2}-m_{c}^{2})\delta(p_{Q}^{2}-m_{Q}^{2})\delta(p_{q}^{2}-m_{q}^{2}).
\label{eq:spectral_density}
\end{eqnarray}
with
\begin{equation}
N_{\mu}={\rm Tr}\left[(\not\!p_{Q}+m_{Q})\gamma^{\rho}(\not\!p_{c}-m_{c})(V_{\mu},A_{\mu})(\not\!p_{b}+m_{b})\gamma^{\nu}\right]\frac{6}{\sqrt{2}}(\gamma_{\rho}\gamma_{5}(\not\!p_{q}+m_{q})\gamma_{5}\gamma_{\nu}).
\end{equation}
The integral in Eq. (\ref{eq:spectral_density}) can be written as a
two-body phase space integral followed by a ``triangle'' phase space
integral \cite{Shi:2019hbf}.

Equating Eq. (\ref{eq:quark}) with Eq. (\ref{eq:hadron}), assuming quark-hadron
duality, and performing the Borel transformation, one can obtain
\begin{equation}
{\cal B}\Pi_{\mu}^{V,{\rm pole}}(T_{1}^{2},T_{2}^{2})=\int^{s_{1}^{0}}ds_{1}\int^{s_{2}^{0}}ds_{2}\ \rho_{\mu}^{V}(s_{1},s_{2},q^{2})e^{-s_{1}/T_{1}^{2}}e^{-s_{2}/T_{2}^{2}},\label{eq:three-point_sum_rule}
\end{equation}
where the left-hand side denotes the
Borel transformed four pole terms in Eq. (\ref{eq:hadron}) and $s_{1,2}^{0}$
are the continuum threshold parameters. Equating the coefficients
of the same Dirac structures on both sides of Eq. (\ref{eq:three-point_sum_rule}),
one can arrive at 12 equations, from which, one can further extract
the form factors. More details can be found in \cite{Shi:2019hbf,Zhao:2020mod}.

In addition, similar as the situation of the two-point correlation function, the LL corrections are also considered.

\section{Numerical results}

The masses of negative-parity baryons are used in this work, and
we adopt the results from \cite{Wang:2010it,Roberts:2007ni}, which are collected in Table~\ref{Tab:mass_negative}.

\begin{table}[htbp!]
\centering \caption{Masses of doubly heavy baryons with $J^{P}=1/2^{-}$ \cite{Wang:2010it,Roberts:2007ni}.}
\begin{tabular}{|l|c|c|c|c|c|c|}
\hline
Baryon  & $\Xi_{cc}(\frac{1}{2}^{-})$  & $\Omega_{cc}(\frac{1}{2}^{-})$  & $\Xi_{bc}(\frac{1}{2}^{-})$  & $\Omega_{bc}(\frac{1}{2}^{-})$  & $\Xi_{bb}(\frac{1}{2}^{-})$  & $\Omega_{bb}(\frac{1}{2}^{-})$\tabularnewline
\hline
Mass  & 3.77  & 3.91  & 7.231  & 7.346  & 10.38  & 10.53\tabularnewline
\hline
\end{tabular}
\label{Tab:mass_negative}
\end{table}

\subsection{Pole residues}

\begin{figure}[htbp!]
   \begin{minipage}[t]{0.3\linewidth}
  \centering
  \includegraphics[width=1.0\columnwidth]{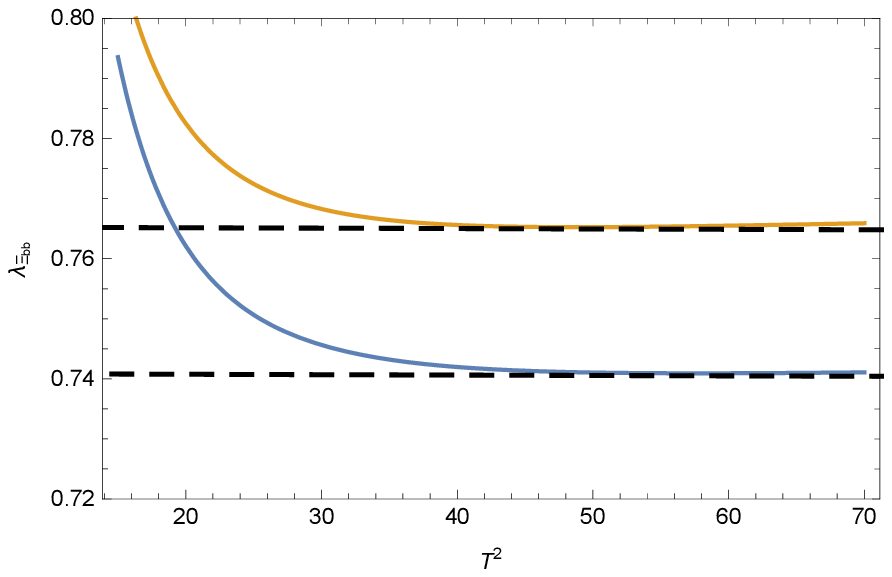}
    \end{minipage}
   \begin{minipage}[t]{0.3\linewidth}
  \centering
  \includegraphics[width=1.0\columnwidth]{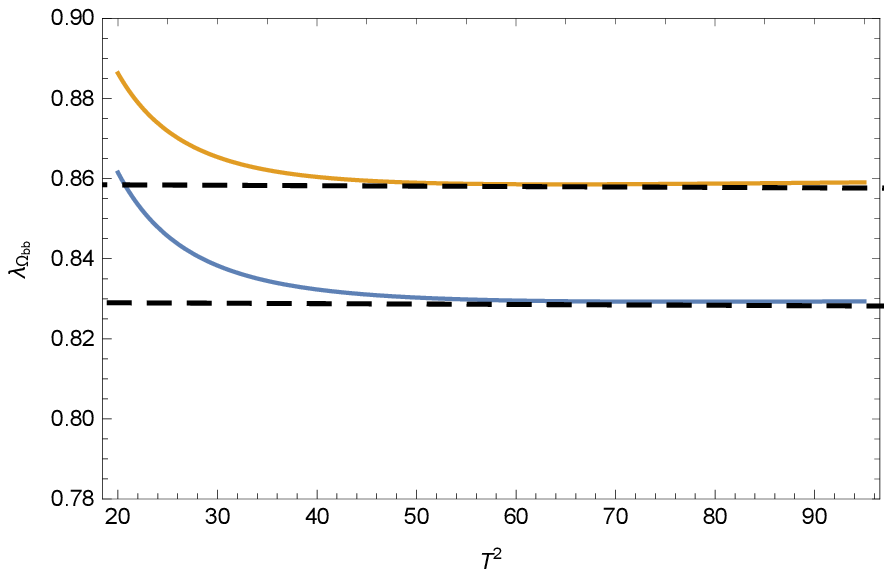}
  \end{minipage}
  \begin{minipage}[t]{0.3\linewidth}
  \centering
  \includegraphics[width=1.0\columnwidth]{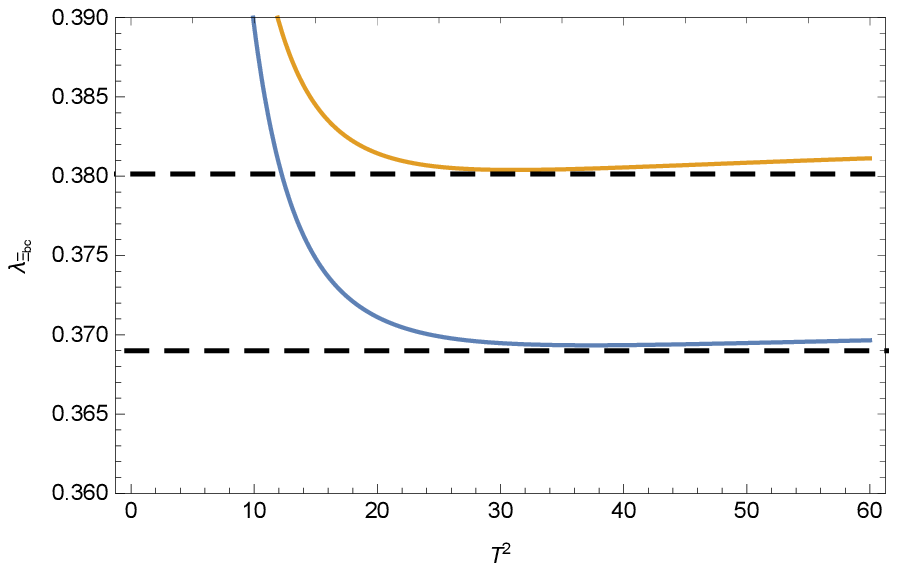}
    \end{minipage}
   \begin{minipage}[t]{0.3\linewidth}
  \centering
  \includegraphics[width=1.0\columnwidth]{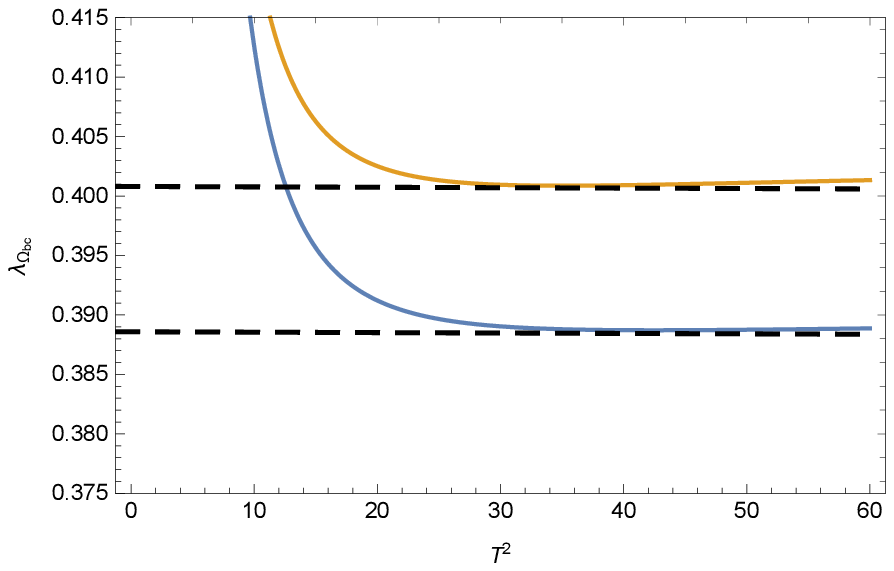}
  \end{minipage}
  \begin{minipage}[t]{0.3\linewidth}
  \centering
  \includegraphics[width=1.0\columnwidth]{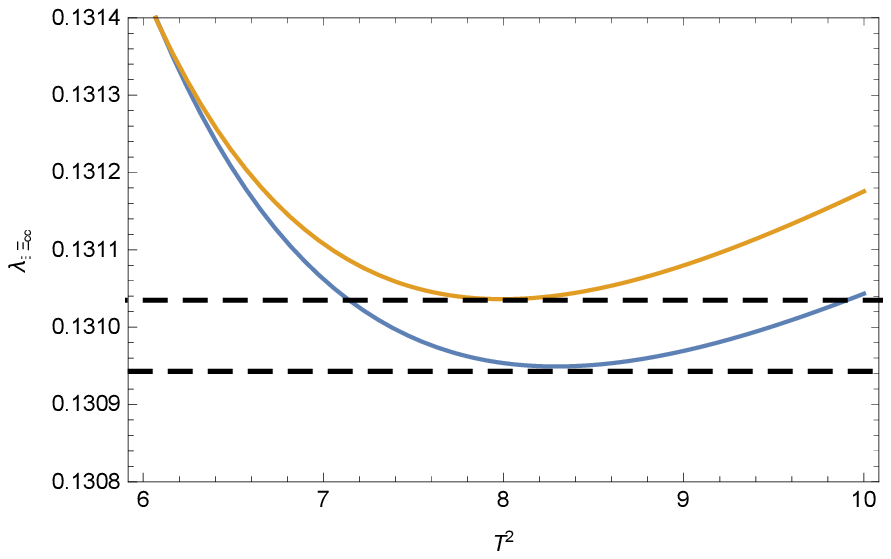}
  \end{minipage}
  \begin{minipage}[t]{0.3\linewidth}
  \centering
  \includegraphics[width=1.0\columnwidth]{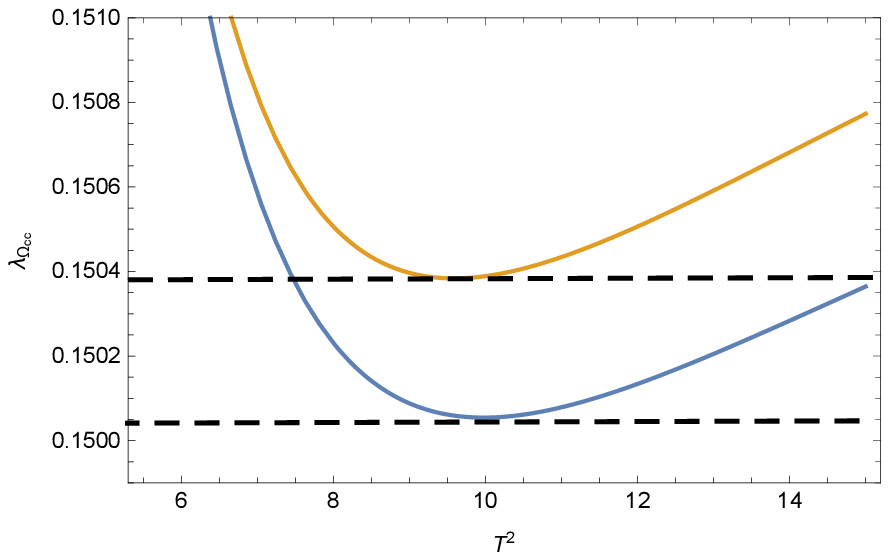}
  \end{minipage}
\caption{The pole residues as functions of the Borel parameters.
The blue and orange curves respectively correspond to our predictions with and without the LL corrections.}
\label{fig:pole_residue}
\end{figure}

Our predictioins of pole residues and masses are respectively collected in Table~\ref{Tab:pole_residue}
and Table~\ref{Tab:mass}, and the pole residues as functions of the Borel parameters are plotted in Fig~\ref{fig:pole_residue}.
Both of the results without and with the LL corrections are shown.
In Table \ref{Tab:mass}, our predictions for the masses are also compared with those
from Lattice QCD \cite{Brown:2014ena}.

\begin{table}[htbp!]
\centering \caption{The pole residues.}
\begin{tabular}{|l|c|c|}
\hline
 & This work (without the LL corrections)  & This work (with the LL corrections)\tabularnewline
\hline
$\lambda_{\Xi_{bb}}$  & 0.760 & $0.736_{-0.000}^{+0.000}(T^{2})_{-0.052}^{+0.053}(s_{0})$ \tabularnewline
\hline
$\lambda_{\Omega_{bb}}$  & 0.854 & $0.825_{-0.000}^{+0.000}(T^{2})_{-0.059}^{+0.060}(s_{0})$ \tabularnewline
\hline
$\lambda_{\Xi_{bc}}$  & 0.379 & $0.369_{-0.000}^{+0.000}(T^{2})_{-0.027}^{+0.028}(s_{0})$ \tabularnewline
\hline
$\lambda_{\Omega_{bc}}$  & 0.400 & $0.388_{-0.000}^{+0.000}(T^{2})_{-0.029}^{+0.030}(s_{0})$ \tabularnewline
\hline
$\lambda_{\Xi_{cc}}$  & 0.130 & $0.130_{-0.000}^{+0.000}(T^{2})_{-0.011}^{+0.011}(s_{0})$ \tabularnewline
\hline
$\lambda_{\Omega_{cc}}$  & 0.150 & $0.149_{-0.000}^{+0.000}(T^{2})_{-0.012}^{+0.013}(s_{0})$ \tabularnewline
\hline
\end{tabular}
\label{Tab:pole_residue}
\end{table}

\begin{table}[htbp!]
\centering \caption{The masses.}
\begin{tabular}{|l|c|c|c|}
\hline
 & This work (without the LL corrections)  & This work (with the LL corrections)  & Lattice QCD \cite{Brown:2014ena}\tabularnewline
\hline
$m_{\Xi_{bb}}$  & 10.166  & $10.152_{-0.009}^{+0.007}(T^{2})_{-0.080}^{+0.079}(s_{0})$  & 10.143\tabularnewline
\hline
$m_{\Omega_{bb}}$  & 10.291  & $10.279_{-0.007}^{+0.006}(T^{2})_{-0.081}^{+0.080}(s_{0})$  & 10.273\tabularnewline
\hline
$m_{\Xi_{bc}}$  & 6.948  & $6.935_{-0.007}^{+0.009}(T^{2})_{-0.081}^{+0.080}(s_{0})$  & 6.943 \tabularnewline
\hline
$m_{\Omega_{bc}}$  & 7.002  & $6.998_{-0.008}^{+0.006}(T^{2})_{-0.081}^{+0.081}(s_{0})$  & 6.998 \tabularnewline
\hline
$m_{\Xi_{cc}}$  & 3.634  & $3.629_{-0.012}^{+0.010}(T^{2})_{-0.079}^{+0.078}(s_{0})$  & 3.621 \cite{LHCb:2017iph}\tabularnewline
\hline
$m_{\Omega_{cc}}$  & 3.747  & $3.743_{-0.011}^{+0.009}(T^{2})_{-0.080}^{+0.079}(s_{0})$  & 3.738 \tabularnewline
\hline
\end{tabular}\label{Tab:mass}
\end{table}


\subsection{Form factors}

\begin{figure}[htbp!]
\begin{minipage}[t]{0.3\linewidth}%
 \centering \includegraphics[width=1\columnwidth]{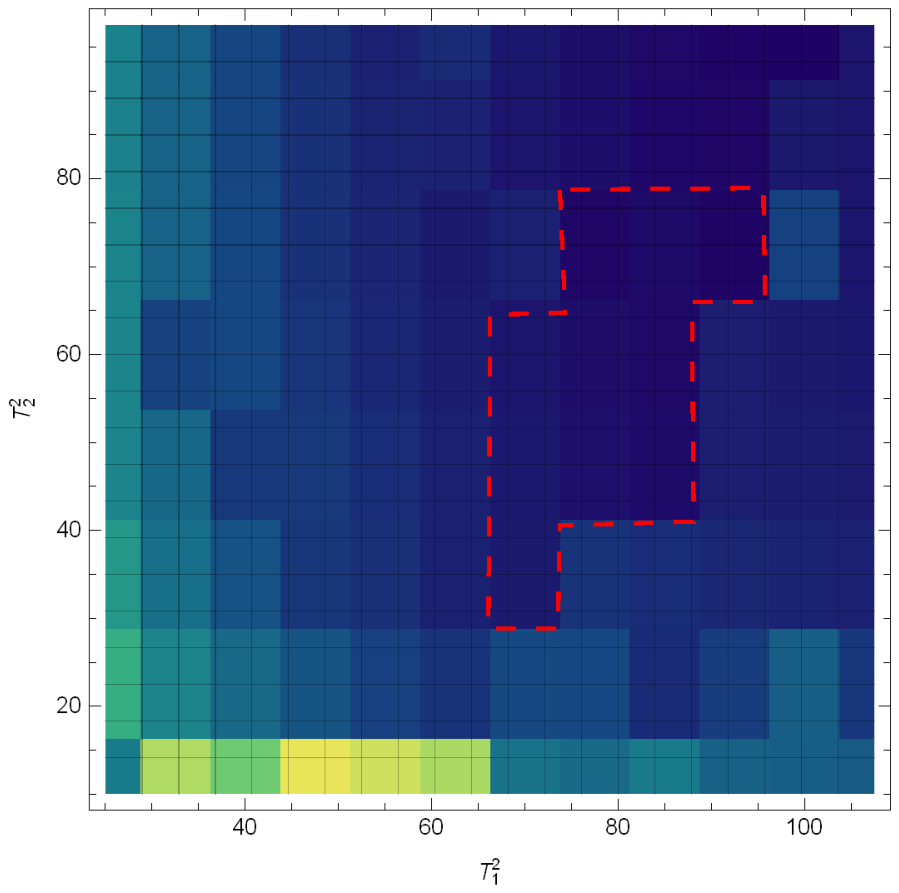} %
\end{minipage}%
\begin{minipage}[t]{0.3\linewidth}%
 \centering \includegraphics[width=1\columnwidth]{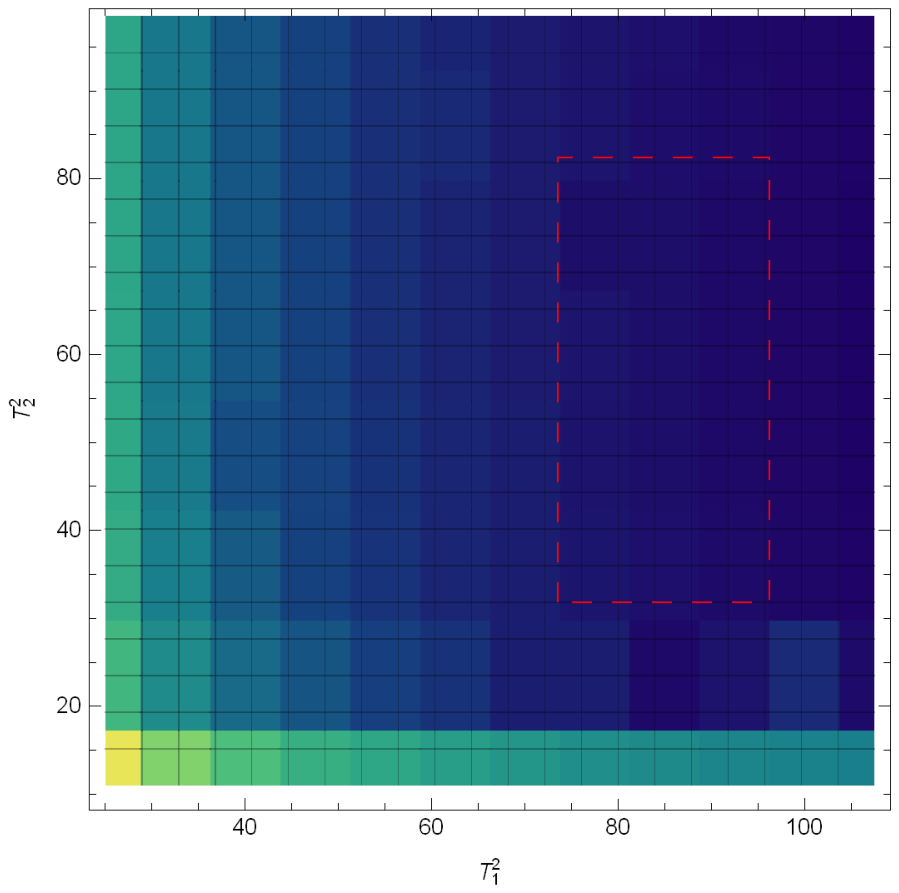} %
\end{minipage}
\begin{minipage}[t]{0.3\linewidth}%
 \centering \includegraphics[width=1\columnwidth]{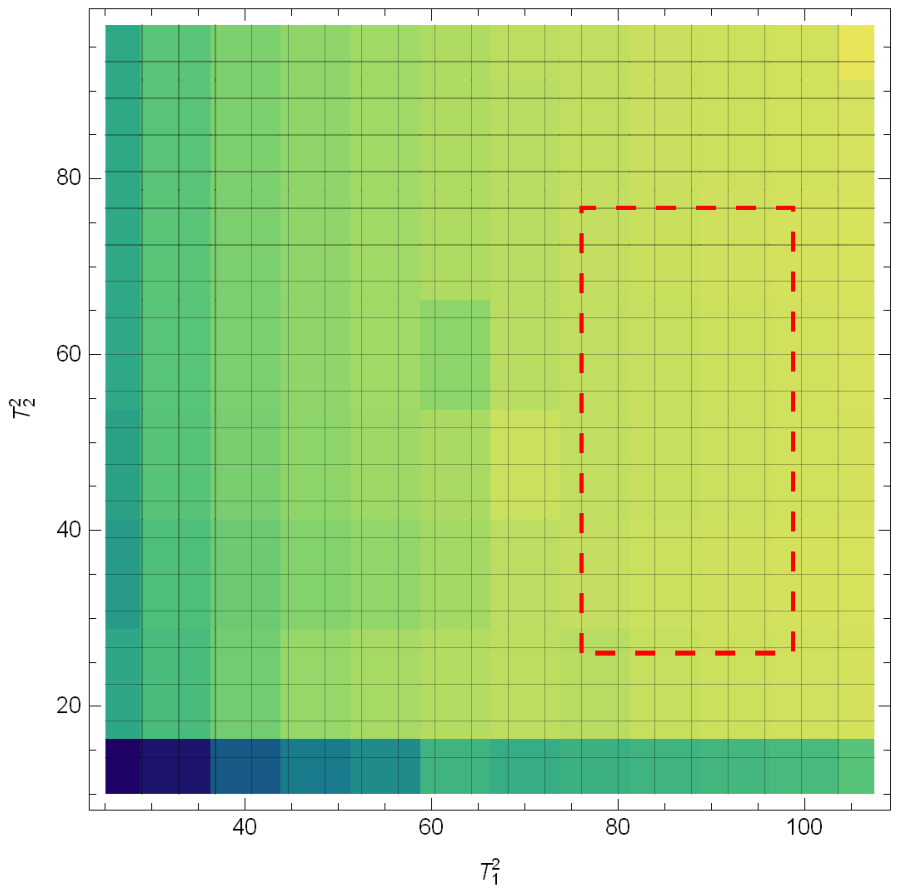} %
\end{minipage}%
\begin{minipage}[t]{0.3\linewidth}%
 \centering \includegraphics[width=1\columnwidth]{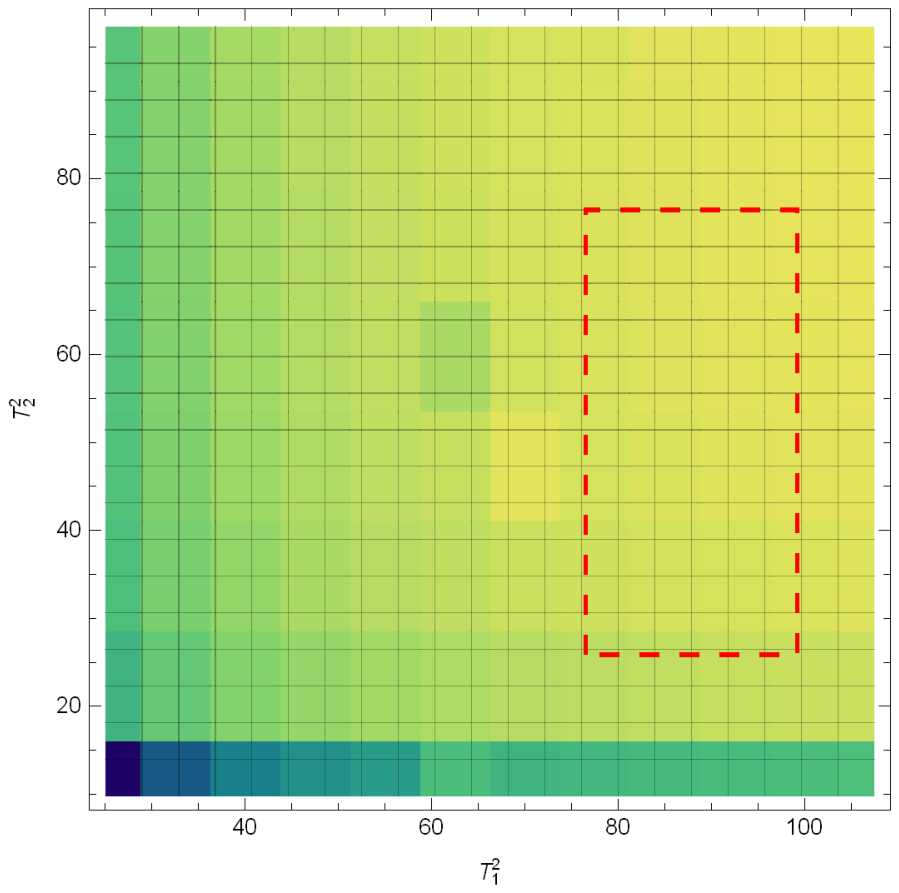} %
\end{minipage}
\caption{The $\Xi_{bb}\to\Xi_{bc}$ form factors $f_{+,0}$ (top left), $f_{\perp}$
(top right), $g_{+,0}$ (bottom left) and $g_{\perp}$ (bottom right) at $q^{2}=0$:
the dependence on the Borel parameters $T_{1,2}^{2}$.}
\label{fig:ff_T1_T2}
\end{figure}

We take the process of $\Xi_{bb}\to\Xi_{bc}$ as an example to illustrate
the selection of Borel windows. In Fig.~\ref{fig:ff_T1_T2}, the transition form factors of $\Xi_{bb}\to\Xi_{bc}$
are plotted as functions of the Borel parameters $T_{1,2}^{2}$. Relatively
flat regions are selected as the working Borel windows.

\begin{figure}[htbp!]
   \begin{minipage}[t]{0.3\linewidth}
  \centering
  \includegraphics[width=1.0\columnwidth]{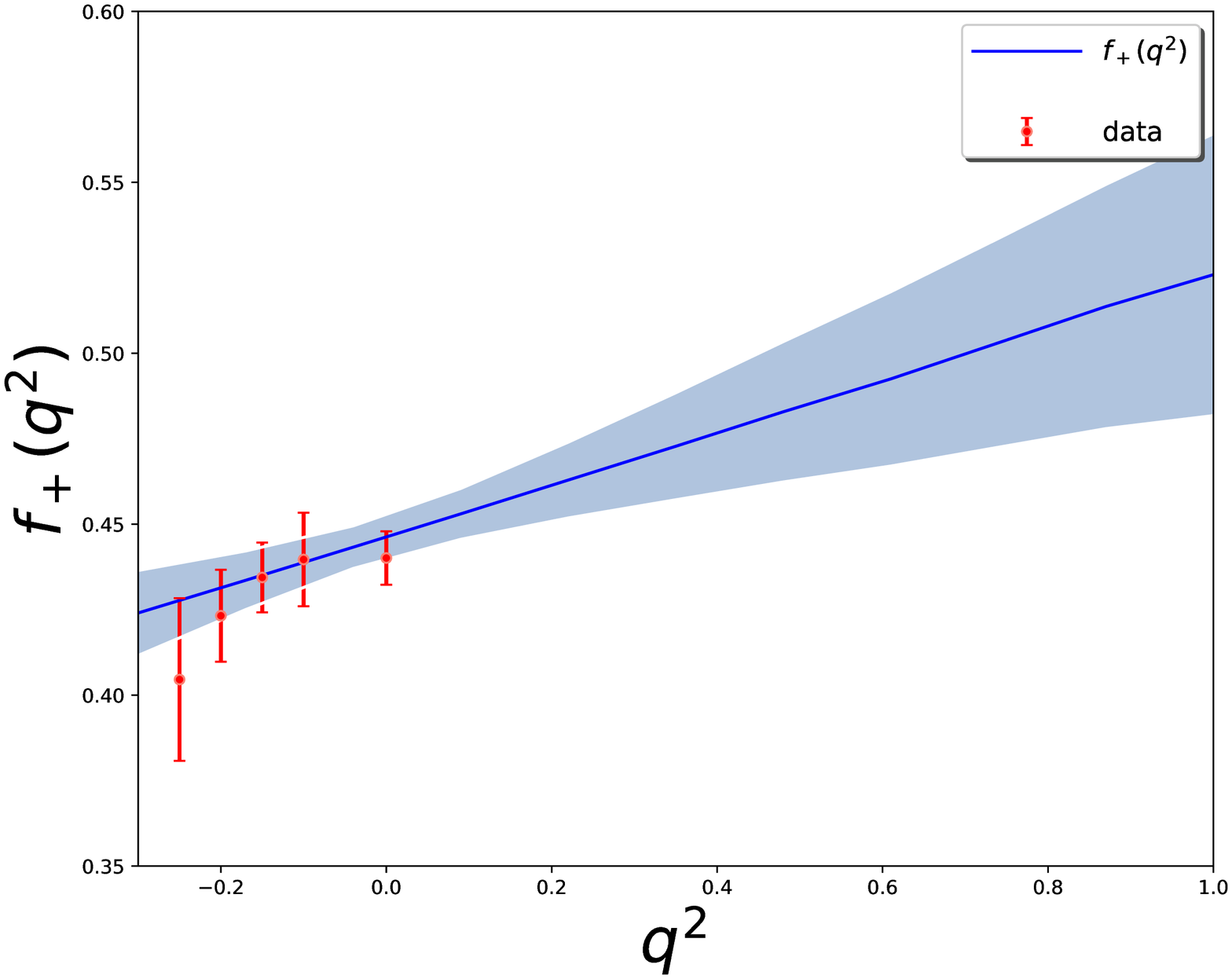}
    \end{minipage}
    \begin{minipage}[t]{0.3\linewidth}
  \centering
  \includegraphics[width=1.0\columnwidth]{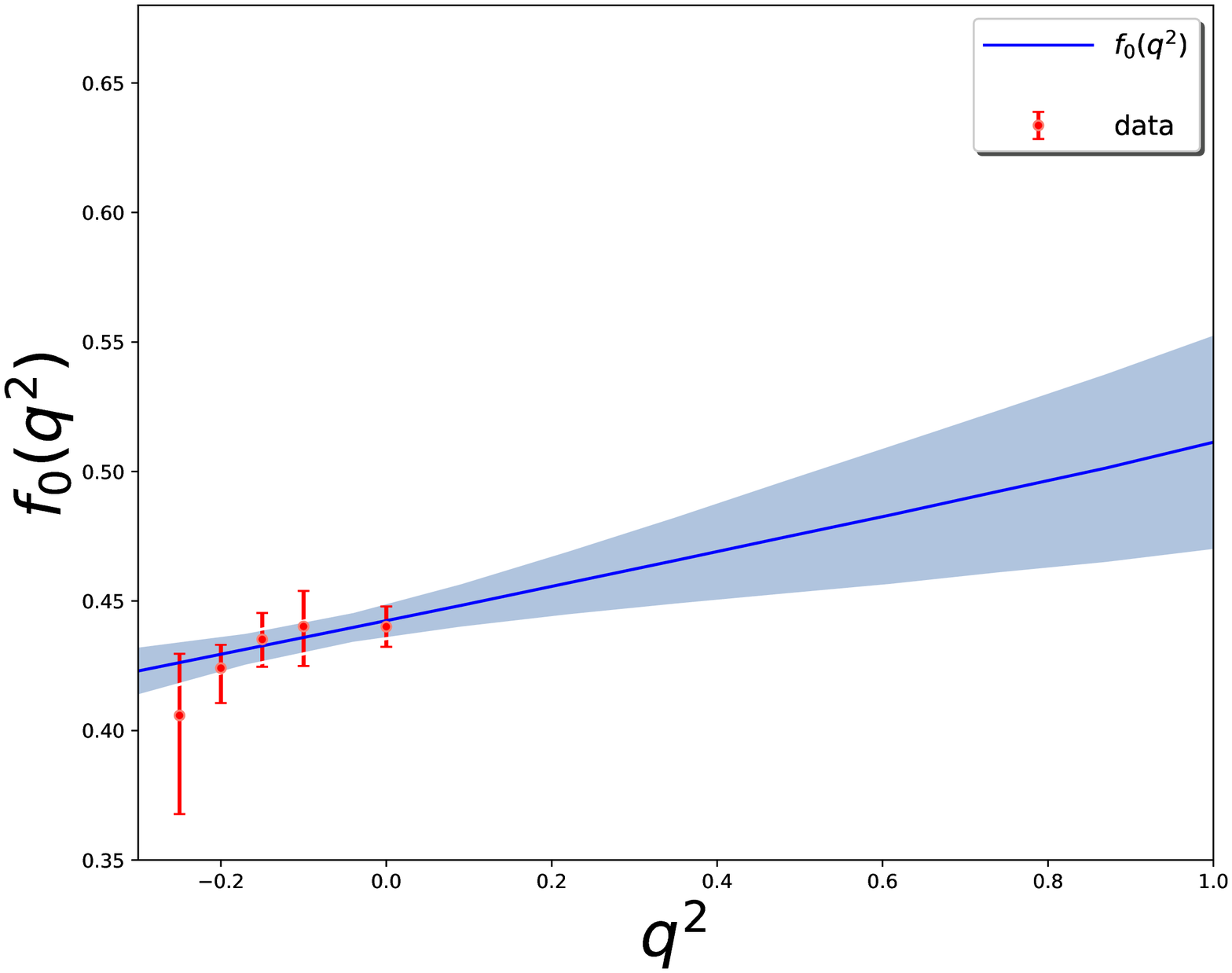}
    \end{minipage}
   \begin{minipage}[t]{0.3\linewidth}
  \centering
  \includegraphics[width=1.0\columnwidth]{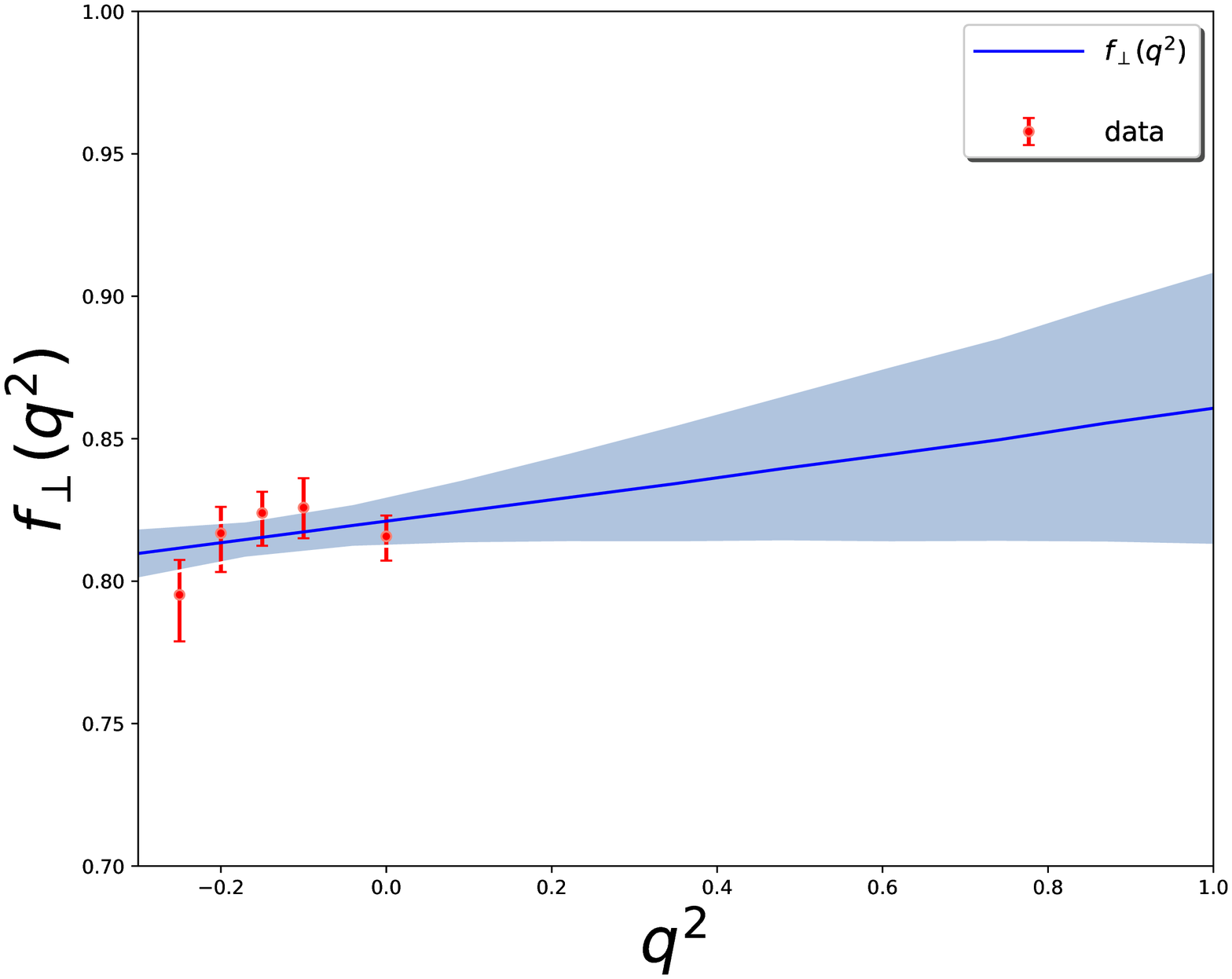}
    \end{minipage}
     \begin{minipage}[t]{0.3\linewidth}
  \centering
  \includegraphics[width=1.0\columnwidth]{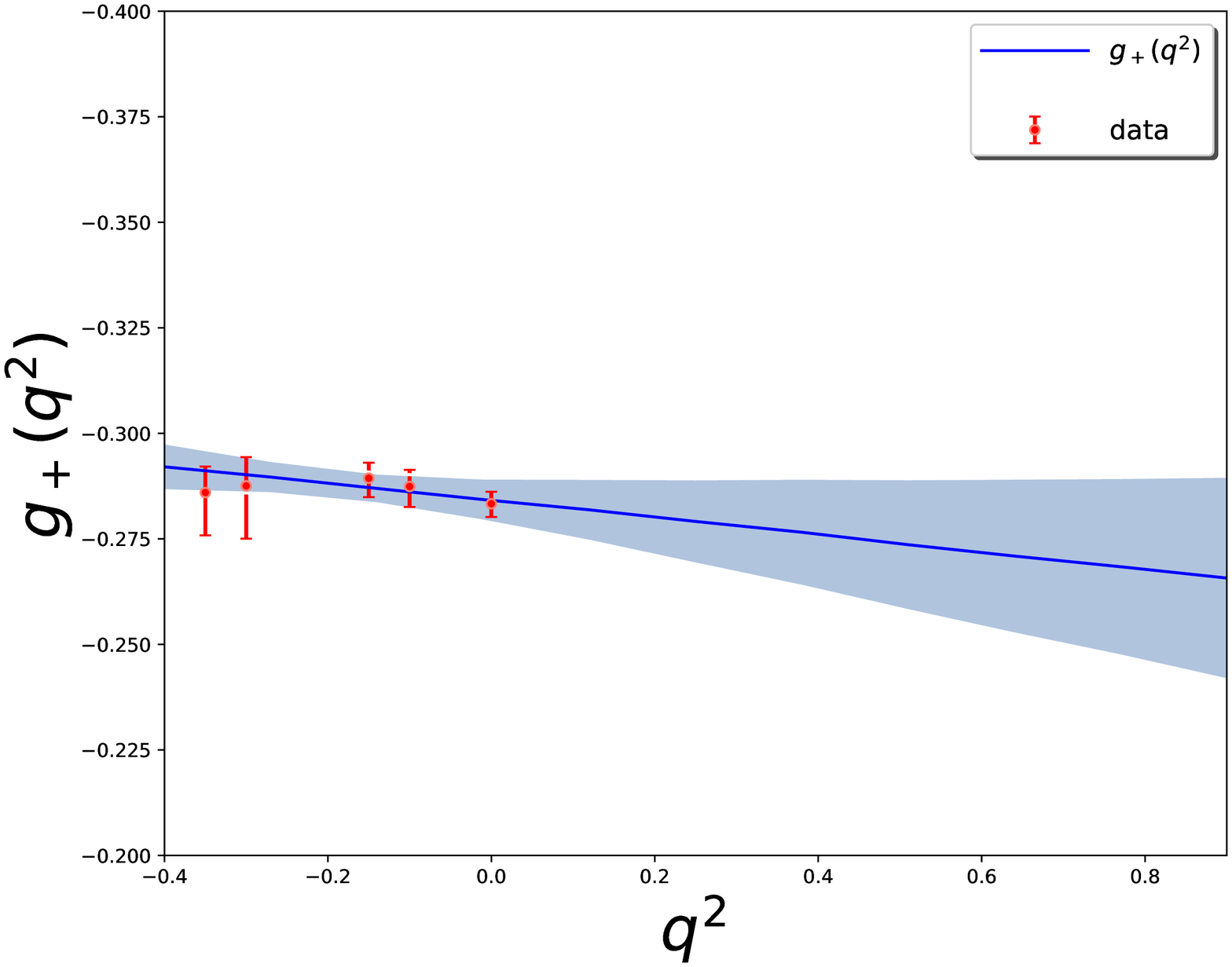}
    \end{minipage}
    \begin{minipage}[t]{0.3\linewidth}
  \centering
  \includegraphics[width=1.0\columnwidth]{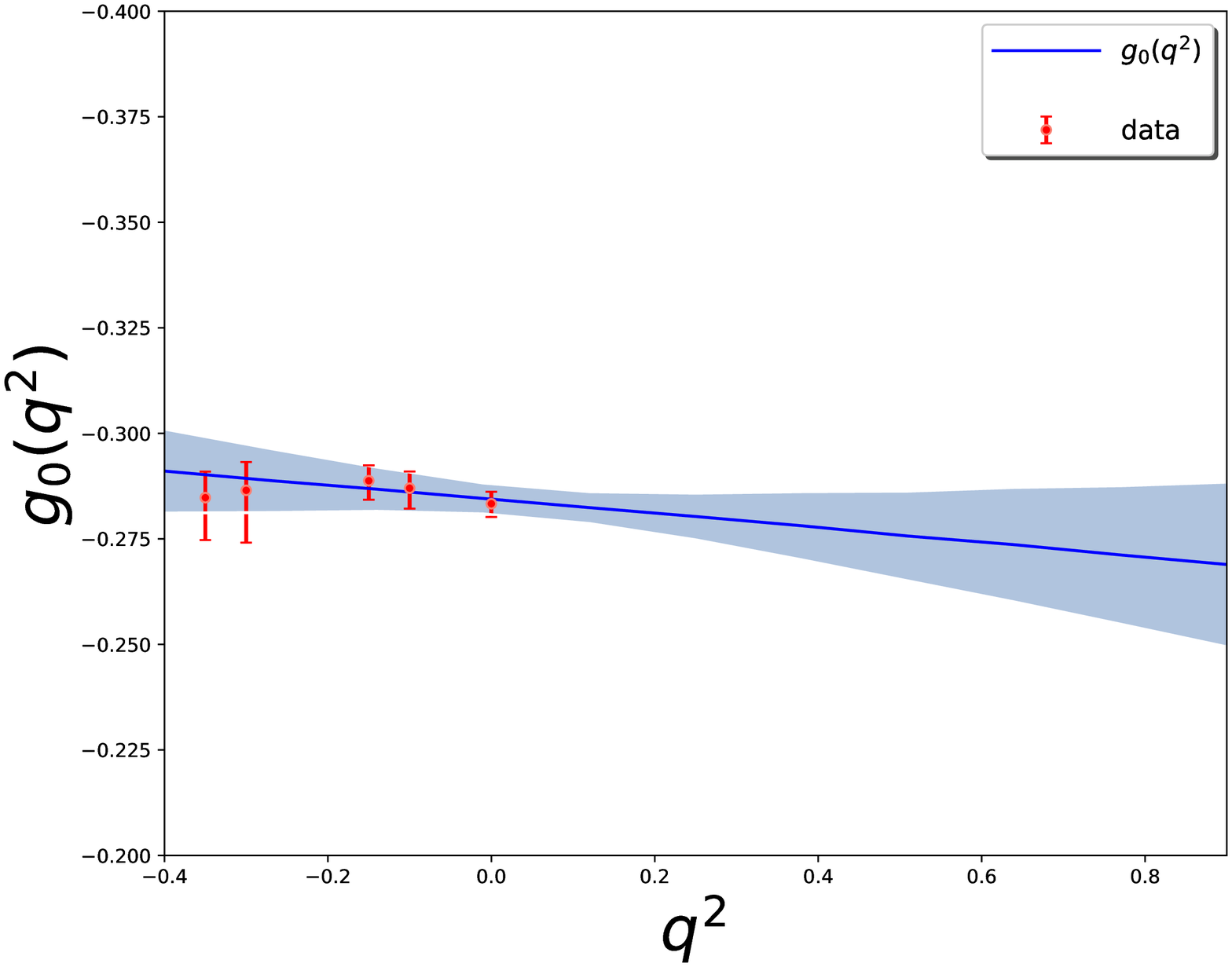}
    \end{minipage}
   \begin{minipage}[t]{0.3\linewidth}
  \centering
  \includegraphics[width=1.0\columnwidth]{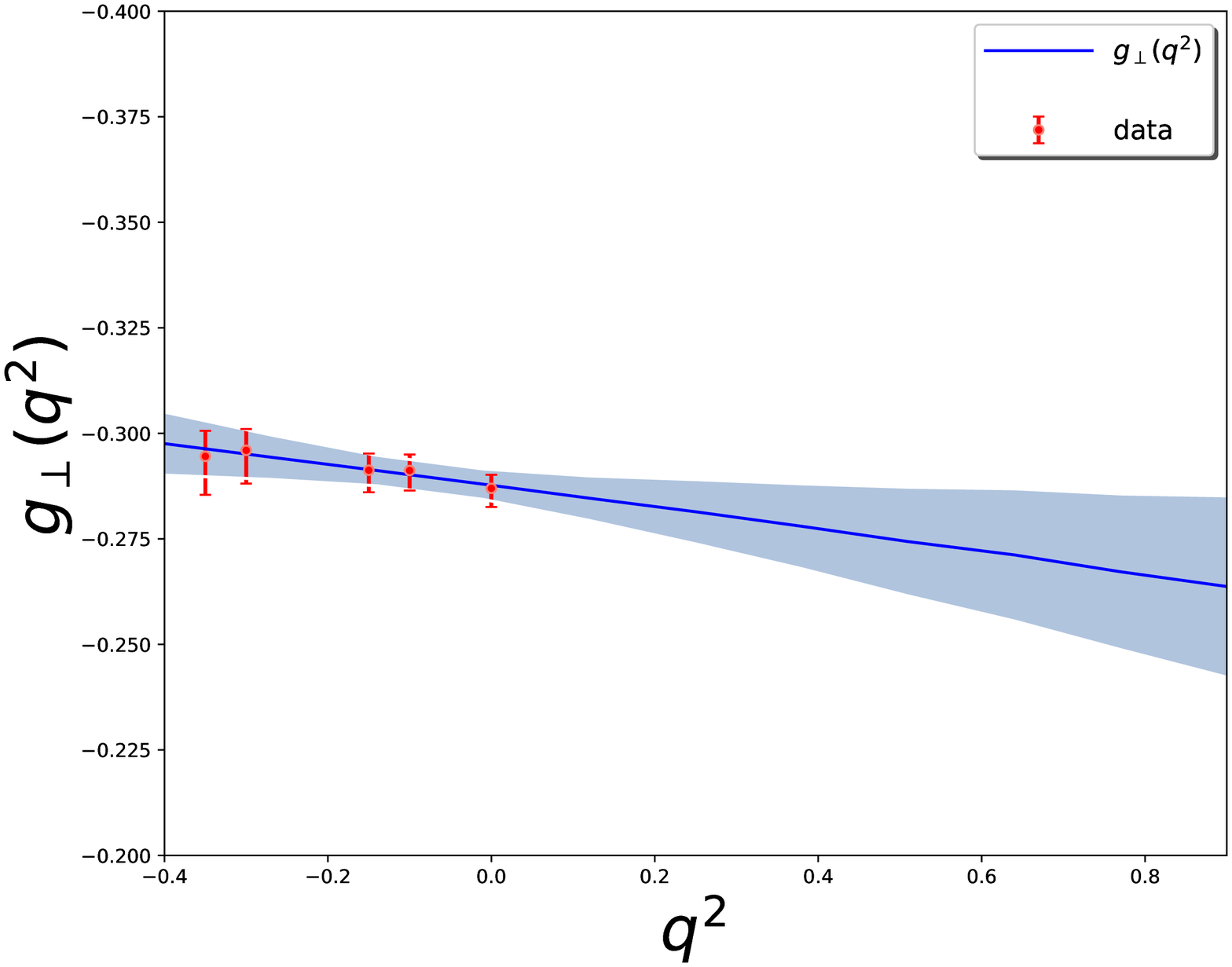}
    \end{minipage}
\caption{The fitted results of the $\Xi_{bb}\to \Xi_{bc}$ form factors.}
\label{fig:ff_q2}
\end{figure}

We also consider the uncertainties of the form factors caused by the
Borel parameters $T_{1,2}^{2}$ and the continuum threshold parameter
$s_{1,2}^{0}$, as can be seen in Table~\ref{Tab:ff}. To access
the $q^{2}$ dependence, we calculate the form factors at small $q^{2}$, and then fit the data with the following formula
\begin{equation}
f(q^{2})=\frac{1}{1-q^{2}/(m_{{\rm pole}})^{2}}(a+bz(q^{2}))
\end{equation}
with
\begin{eqnarray}
z(q^{2})=\frac{\sqrt{t_{+}-q^{2}}-\sqrt{t_{+}-t_{0}}}{\sqrt{t_{+}-q^{2}}+\sqrt{t_{+}-t_{0}}}.
\end{eqnarray}
Here $t_{0}=q_{{\rm max}}^{2}=(m_{\mathcal{B}_{i}}-m_{\mathcal{B}_{f}})^{2}$
and $t_{+}=(m_{{\rm pole}})^{2}$ are chosen to be equal or below
the location of any remaining singularity after factoring out the
leading pole contribution \cite{Detmold:2015aaa}. The nonlinear least-$\chi^2$ (lsq)
method is used in our analysis~\cite{Peter:2020}. The fitted results
are shown in Table~\ref{Tab:ff} and Fig.~\ref{fig:ff_q2}. The contributions of each local operator
in the OPE are evaluated as
\begin{eqnarray}
f_{+,0}^{\Xi_{bb}\to\Xi_{bc}}(0) & = & 0.332(73.309\%)_{dim0}+0.101(22.300\%)_{dim3}+0.014(3.063\%)_{dim5}-0.006(1.329\%)_{dim6},\nonumber \\
f_{\perp}^{\Xi_{bb}\to\Xi_{bc}}(0) & = & 0.630(75.700\%)_{dim0}+0.194(23.326\%)_{dim3}-0.003(0.310\%)_{dim5}-0.006(0.663\%)_{dim6}.
\end{eqnarray}
It can be seen that the OPE has excellent convergence and the perturbative
term dominates.

In addition, we have investigated the heavy quark limit of our full
QCD results for the form factors, and close results are obtained.

\begin{table}[htbp!]
\centering \caption{The fitted results for the form factors. The parameter $a$ and $b$ satisfy a two-dimensional Gaussian distribution with $\rho=-0.999$.}
\begin{tabular}{|l|c|c|c|}
\hline
 & $F(0)$  & $a$  & $b$\tabularnewline
\hline
$f_{+}^{\Xi_{bb}\to\Xi_{bc}}$  & $ 0.441_{-0.008}^{+0.008}(T_{1}^{2},T_{2}^{2})_{-0.015}^{+0.020}(s_{1}^0)_{-0.037}^{+0.072}(s_{2}^0)$  & $1.200\pm0.460(T_{1}^{2},T_{2}^{2})\pm0.600(s_{1}^0,s_{2}^0)$ & $-10.1\pm6.1(T_{1}^{2},T_{2}^{2})\pm7.9(s_{1}^0,s_{2}^0)$\tabularnewline
\hline
$f_{0}^{\Xi_{bb}\to\Xi_{bc}}$  & $ 0.441_{-0.008}^{+0.008}(T_{1}^{2},T_{2}^{2})_{-0.015}^{+0.020}(s_{1}^0)_{-0.037}^{+0.072}(s_{2}^0)$  & $1.090\pm0.440(T_{1}^{2},T_{2}^{2})\pm0.600(s_{1}^0,s_{2}^0)$ & $-8.6\pm5.8(T_{1}^{2},T_{2}^{2})\pm7.9(s_{1}^0,s_{2}^0)$\tabularnewline
\hline
$f_{\perp}^{\Xi_{bb}\to\Xi_{bc}}$  & $0.816_{-0.007}^{+0.009}(T_{1}^{2},T_{2}^{2})_{-0.047}^{+0.045}(s_{1}^0)_{-0.092}^{+0.114}(s_{2}^0)$  & $1.030\pm0.480(T_{1}^{2},T_{2}^{2})\pm0.710(s_{1}^0,s_{2}^0)$  & $-2.7\pm6.3(T_{1}^{2},T_{2}^{2})\pm9.3(s_{1}^0,s_{2}^0)$ \tabularnewline
\hline
$g_{+}^{\Xi_{bb}\to\Xi_{bc}}$  & $-0.283_{-0.003}^{+0.003}(T_{1}^{2},T_{2}^{2})_{-0.027}^{+0.033}(s_{1}^0)_{-0.050}^{+0.048}(s_{2}^0)$  & $0.040\pm0.26(T_{1}^{2},T_{2}^{2})\pm0.710(s_{1}^0,s_{2}^0)$  & $-4.3\pm3.4(T_{1}^{2},T_{2}^{2})\pm9.3(s_{1}^0,s_{2}^0)$\tabularnewline
\hline
$g_{0}^{\Xi_{bb}\to\Xi_{bc}}$  & $-0.283_{-0.003}^{+0.003}(T_{1}^{2},T_{2}^{2})_{-0.027}^{+0.033}(s_{1}^0)_{-0.050}^{+0.048}(s_{2}^0)$  & $-0.001\pm0.255(T_{1}^{2},T_{2}^{2})\pm0.710(s_{1}^0,s_{2}^0)$  & $-3.8\pm3.4(T_{1}^{2},T_{2}^{2})\pm9.3(s_{1}^0,s_{2}^0)$\tabularnewline
\hline
$g_{\perp}^{\Xi_{bb}\to\Xi_{bc}}$  & $-0.287^{+0.003}_{-0.004}(T_{1}^{2},T_{2}^{2})_{-0.016}^{+0.021}(s_{1}^0)_{-0.040}^{+0.038}(s_{2}^0)$  & $0.100\pm0.250(T_{1}^{2},T_{2}^{2})\pm0.590(s_{1}^0,s_{2}^0)$ & $-5.1\pm3.3(T_{1}^{2},T_{2}^{2})\pm7.8(s_{1}^0,s_{2}^0)$\tabularnewline
\hline
$f_{+}^{\Omega_{bb}\to\Omega_{bc}}$  & $0.463_{-0.008}^{+0.004}(T_{1}^{2},T_{2}^{2})_{-0.047}^{+0.041}(s_{1}^0)_{-0.070}^{+0.056}(s_{2}^0)$  & $0.310\pm0.480(T_{1}^{2},T_{2}^{2})\pm0.740(s_{1}^0,s_{2}^0)$ & $1.8\pm6.0(T_{1}^{2},T_{2}^{2})\pm9.2(s_{1}^0,s_{2}^0)$\tabularnewline
\hline
$f_{0}^{\Omega_{bb}\to\Omega_{bc}}$  & $0.462_{-0.008}^{+0.004}(T_{1}^{2},T_{2}^{2})_{-0.047}^{+0.041}(s_{1}^0)_{-0.070}^{+0.056}(s_{2}^0)$  & $0.320\pm0.480(T_{1}^{2},T_{2}^{2})\pm0.740(s_{1}^0,s_{2}^0)$ & $1.7\pm5.9(T_{1}^{2},T_{2}^{2})\pm9.2(s_{1}^0,s_{2}^0)$\tabularnewline
\hline
$f_{\perp}^{\Omega_{bb}\to\Omega_{bc}}$  & $0.865_{-0.009}^{+0.010}(T_{1}^{2},T_{2}^{2})_{-0.079}^{+0.013}(s_{1}^0)_{-0.124}^{+0.082}(s_{2}^0)$  & $1.646\pm0.600(T_{1}^{2},T_{2}^{2})\pm2.500(s_{1}^0,s_{2}^0)$  & $-9.8\pm7.4(T_{1}^{2},T_{2}^{2})\pm31(s_{1}^0,s_{2}^0)$ \tabularnewline
\hline
$g_{+}^{\Omega_{bb}\to\Omega_{bc}}$  & $-0.286_{-0.003}^{+0.003}(T_{1}^{2},T_{2}^{2})_{-0.057}^{+0.035}(s_{1}^0)_{-0.065}^{+0.065}(s_{2}^0)$  & $-0.560\pm0.170(T_{1}^{2},T_{2}^{2})\pm0.740(s_{1}^0,s_{2}^0)$  & $3.4\pm2.1(T_{1}^{2},T_{2}^{2})\pm9.2(s_{1}^0,s_{2}^0)$\tabularnewline
\hline
$g_{0}^{\Omega_{bb}\to\Omega_{bc}}$  & $-0.286_{-0.003}^{+0.003}(T_{1}^{2},T_{2}^{2})_{-0.057}^{+0.035}(s_{1}^0)_{-0.065}^{+0.065}(s_{2}^0)$  & $-0.600\pm0.170(T_{1}^{2},T_{2}^{2})\pm0.740(s_{1}^0,s_{2}^0)$  & $4.0\pm2.1(T_{1}^{2},T_{2}^{2})\pm9.2(s_{1}^0,s_{2}^0)$\tabularnewline
\hline
$g_{\perp}^{\Omega_{bb}\to\Omega_{bc}}$  & $-0.292_{-0.002}^{+0.002}(T_{1}^{2},T_{2}^{2})_{-0.037}^{+0.019}(s_{1}^0)_{-0.042}^{+0.044}(s_{2}^0)$  & $-0.510\pm0.130(T_{1}^{2},T_{2}^{2})\pm2.100(s_{1}^0,s_{2}^0)$ & $2.8\pm1.7(T_{1}^{2},T_{2}^{2})\pm26(s_{1}^0,s_{2}^0)$\tabularnewline
\hline
$f_{+}^{\Xi_{bc}\to\Xi_{cc}}$  & $0.671_{-0.004}^{+0.007}(T_{1}^{2},T_{2}^{2})_{-0.052}^{+0.050}(s_{1}^0)_{-0.137}^{+0.142}(s_{2}^0)$  & $1.380\pm0.330(T_{1}^{2},T_{2}^{2})\pm1.500(s_{1}^0,s_{2}^0)$ & $-8.5\pm3.9(T_{1}^{2},T_{2}^{2})\pm18(s_{1}^0,s_{2}^0)$\tabularnewline
\hline
$f_{0}^{\Xi_{bc}\to\Xi_{cc}}$  & $0.671_{-0.004}^{+0.007}(T_{1}^{2},T_{2}^{2})_{-0.052}^{+0.050}(s_{1}^0)_{-0.137}^{+0.142}(s_{2}^0)$  & $1.310\pm0.330(T_{1}^{2},T_{2}^{2})\pm2.000(s_{1}^0,s_{2}^0)$ & $-7.7\pm4.0(T_{1}^{2},T_{2}^{2})\pm24(s_{1}^0,s_{2}^0)$\tabularnewline
\hline
$f_{\perp}^{\Xi_{bc}\to\Xi_{cc}}$  & $0.815_{-0.005}^{+0.006}(T_{1}^{2},T_{2}^{2})_{-0.063}^{+0.061}(s_{1}^0)_{-0.169}^{+0.175}(s_{2}^0)$  & $1.340\pm0.320(T_{1}^{2},T_{2}^{2})\pm3.000(s_{1}^0,s_{2}^0)$  & $-6.3\pm3.9(T_{1}^{2},T_{2}^{2})\pm36(s_{1}^0,s_{2}^0)$ \tabularnewline
\hline
$g_{+}^{\Xi_{bc}\to\Xi_{cc}}$  & $-0.435_{-0.002}^{+0.002}(T_{1}^{2},T_{2}^{2})_{-0.083}^{+0.82}(s_{1}^0)_{-0.123}^{+0.121}(s_{2}^0)$  & $-0.310\pm0.130(T_{1}^{2},T_{2}^{2})\pm0.820(s_{1}^0,s_{2}^0)$  & $-1.6\pm1.6(T_{1}^{2},T_{2}^{2})\pm9.8(s_{1}^0,s_{2}^0)$\tabularnewline
\hline
$g_{0}^{\Xi_{bc}\to\Xi_{cc}}$  & $-0.435_{-0.002}^{+0.002}(T_{1}^{2},T_{2}^{2})_{-0.083}^{+0.077}(s_{1}^0)_{-0.123}^{+0.121}(s_{2}^0)$  & $-0.370\pm0.150(T_{1}^{2},T_{2}^{2})\pm0.770(s_{1}^0,s_{2}^0)$  & $-0.8\pm1.8(T_{1}^{2},T_{2}^{2})\pm9.2(s_{1}^0,s_{2}^0)$\tabularnewline
\hline
$g_{\perp}^{\Xi_{bc}\to\Xi_{cc}}$  & $-0.456_{-0.002}^{+0.002}(T_{1}^{2},T_{2}^{2})_{-0.046}^{+0.040}(s_{1}^0)_{-0.080}^{+0.075}(s_{2}^0)$  & $-0.370\pm0.130(T_{1}^{2},T_{2}^{2})\pm0.760(s_{1}^0,s_{2}^0)$  & $-1.1\pm1.6(T_{1}^{2},T_{2}^{2})\pm9.2(s_{1}^0,s_{2}^0)$\tabularnewline
\hline
$f_{+}^{\Omega_{bc}\to\Omega_{cc}}$  & $0.659_{-0.000}^{+0.001}(T_{1}^{2},T_{2}^{2})_{-0.064}^{+0.050}(s_{1}^0)_{-0.137}^{+0.142}(s_{2}^0)$  & $0.960\pm0.130(T_{1}^{2},T_{2}^{2})\pm1.400(s_{1}^0,s_{2}^0)$ & $-3.8\pm1.6(T_{1}^{2},T_{2}^{2})\pm18(s_{1}^0,s_{2}^0)$\tabularnewline
\hline
$f_{0}^{\Omega_{bc}\to\Omega_{cc}}$  & $0.659_{-0.000}^{+0.001}(T_{1}^{2},T_{2}^{2})_{-0.064}^{+0.050}(s_{1}^0)_{-0.137}^{+0.142}(s_{2}^0)$  & $0.890\pm0.130(T_{1}^{2},T_{2}^{2})\pm1.400(s_{1}^0,s_{2}^0)$ & $-2.9\pm1.6(T_{1}^{2},T_{2}^{2})\pm18(s_{1}^0,s_{2}^0)$\tabularnewline
\hline
$f_{\perp}^{\Omega_{bc}\to\Omega_{cc}}$  & $0.804_{-0.001}^{+0.001}(T_{1}^{2},T_{2}^{2})_{-0.077}^{+0.058}(s_{1}^0)_{-0.169}^{+0.175}(s_{2}^0)$  & $1.510\pm0.170(T_{1}^{2},T_{2}^{2})\pm2.400(s_{1}^0,s_{2}^0)$  & $-9.0\pm2.1(T_{1}^{2},T_{2}^{2})\pm34(s_{1}^0,s_{2}^0)$ \tabularnewline
\hline
$g_{+}^{\Omega_{bc}\to\Omega_{cc}}$  & $-0.410_{-0.003}^{+0.003}(T_{1}^{2},T_{2}^{2})_{-0.079}^{+0.087}(s_{1}^0)_{-0.293}^{+0.270}(s_{2}^0)$  & $-0.630\pm0.280(T_{1}^{2},T_{2}^{2})\pm0.750(s_{1}^0,s_{2}^0)$  & $2.8\pm3.5(T_{1}^{2},T_{2}^{2})\pm9.3(s_{1}^0,s_{2}^0)$\tabularnewline
\hline
$g_{0}^{\Omega_{bc}\to\Omega_{cc}}$  & $-0.410_{-0.003}^{+0.003}(T_{1}^{2},T_{2}^{2})_{-0.079}^{+0.087}(s_{1}^0)_{-0.293}^{+0.270}(s_{2}^0)$  & $-0.960\pm0.130(T_{1}^{2},T_{2}^{2})\pm0.750(s_{1}^0,s_{2}^0)$  & $-3.8\pm1.6(T_{1}^{2},T_{2}^{2})\pm9.3(s_{1}^0,s_{2}^0)$\tabularnewline
\hline
$g_{\perp}^{\Omega_{bc}\to\Omega_{cc}}$  & $-0.429_{-0.002}^{+0.002}(T_{1}^{2},T_{2}^{2})_{-0.042}^{+0.049}(s_{1}^0)_{-0.190}^{+0.172}(s_{2}^0)$  & $-0.810\pm0.250(T_{1}^{2},T_{2}^{2})\pm0.790(s_{1}^0,s_{2}^0)$  & $4.8\pm3.2(T_{1}^{2},T_{2}^{2})\pm9.8(s_{1}^0,s_{2}^0)$\tabularnewline
\hline
\end{tabular}
\label{Tab:ff}
\end{table}

\section{Phenomenological applications}

The weak decays of doubly heavy baryons induced by $b\to c\ l^{-}\bar{\nu}$
can be calculated using the low energy effective Hamiltonian
\begin{eqnarray}
{\cal H}_{{\rm eff}}(b\to c~l^{-}\bar{\nu}_{l}) & = & \frac{G_{F}}{\sqrt{2}}V_{cb}[\bar{c}\gamma^{\mu}(1-\gamma_{5})b][\bar{l}\gamma_{\mu}(1-\gamma_{5})\nu_{l}].\label{eq:efhvb}
\end{eqnarray}
The helicity amplitudes are defined as follows:
\begin{eqnarray}
 & HV_{\lambda^{\prime},\lambda_{W}}^{\lambda} & =\langle{\cal B}_{f}(\lambda^{\prime})|\bar{c}\gamma^{\mu}b|{\cal B}_{i}(\lambda)\rangle\epsilon_{\mu}^{*}(\lambda_{W}),\nonumber \\
 & HA_{\lambda^{\prime},\lambda_{W}}^{\lambda} & =\langle{\cal B}_{f}(\lambda^{\prime})|\bar{c}\gamma^{\mu}\gamma_{5}b|{\cal B}_{i}(\lambda)\rangle\epsilon_{\mu}^{*}(\lambda_{W}),\nonumber \\
 & H_{\lambda^{\prime},\lambda_{W}}^{\lambda}= & HV_{\lambda^{\prime},\lambda_{W}}^{\lambda}-HA_{\lambda^{\prime},\lambda_{W}}^{\lambda}.
\end{eqnarray}
The differential decay widths can be shown as:
\begin{eqnarray}
\frac{d\Gamma}{dq^{2}} & = & \frac{d\Gamma_{L}}{dq^{2}}+\frac{d\Gamma_{T}}{dq^{2}},\nonumber \\
\frac{d\Gamma_{T}}{dq^{2}} & = & \frac{G_{F}^{2}|V_{cb}|^{2}|P^{\prime}||p_{1}|}{16(2\pi)^{3}M_{{\cal B}}^{2}\sqrt{q^{2}}}\frac{2(m_{l}^{2}-q^{2})(m_{l}^{2}+2q^{2})}{3q^{2}}\bigg(\lvert H_{\frac{1}{2},1}^{\frac{1}{2}}\rvert^{2}+\lvert H_{-\frac{1}{2},-1}^{-\frac{1}{2}}\rvert^{2}\bigg),\nonumber \\
\frac{d\Gamma_{L}}{dq^{2}} & = & \frac{G_{F}^{2}|V_{cb}|^{2}|P^{\prime}||p_{1}|}{16(2\pi)^{3}M_{{\cal B}}^{2}\sqrt{q^{2}}}\frac{2(m_{l}^{2}-q^{2})}{3q^{2}}\bigg((m_{l}^{2}+2q^{2})(\lvert H_{-\frac{1}{2},0}^{\frac{1}{2}}\rvert^{2}+\lvert H_{\frac{1}{2},0}^{-\frac{1}{2}}\rvert^{2})+3m_{l}^{2}(\lvert H_{-\frac{1}{2},t}^{\frac{1}{2}}\rvert^{2}+\lvert H_{\frac{1}{2},t}^{-\frac{1}{2}}\rvert^{2})\bigg).\label{eq:dGamma}
\end{eqnarray}
Here $|P^{\prime}|$ is the magnitude of three-momentum of ${\cal B}_{f}$
in the rest frame of ${\cal B}_{i}$, and $|p_{1}|$ is that of lepton in the rest frame of $W$ boson. The helicity amplitudes
in Eq. (\ref{eq:dGamma}) are related to the form factors as follows:
\begin{eqnarray}
HV_{-\frac{1}{2},0}^{\frac{1}{2}} & = & HV_{\frac{1}{2},0}^{-\frac{1}{2}}=-if_{+}(M_{1}+M_{2})\sqrt{\frac{Q_{-}}{q^{2}}},\nonumber \\
HV_{-\frac{1}{2},t}^{\frac{1}{2}} & = & HV_{\frac{1}{2},t}^{-\frac{1}{2}}=-if_{0}(M_{1}-M_{2})\sqrt{\frac{Q_{+}}{q^{2}}},\nonumber \\
HV_{\frac{1}{2},1}^{\frac{1}{2}} & = & HV_{-\frac{1}{2},-1}^{-\frac{1}{2}}=-if_{\perp}\sqrt{2Q_{-}},
\end{eqnarray}
and
\begin{eqnarray}
HA_{-\frac{1}{2},0}^{\frac{1}{2}} & = & -HA_{\frac{1}{2},0}^{-\frac{1}{2}}=ig_{+}(M_{1}-M_{2})\sqrt{\frac{Q_{+}}{q^{2}}},\nonumber \\
HA_{-\frac{1}{2},t}^{\frac{1}{2}} & = & -HA_{\frac{1}{2},t}^{-\frac{1}{2}}=ig_{0}(M_{1}+M_{2})\sqrt{\frac{Q_{-}}{q^{2}}},\nonumber \\
HA_{\frac{1}{2},1}^{\frac{1}{2}} & = & -HA_{-\frac{1}{2},-1}^{-\frac{1}{2}}=-ig_{\perp}\sqrt{2Q_{+}}.
\end{eqnarray}

Our predictions of the decay widths are given in Table~\ref{tab:ccq_semi}.
 \begin{table}[htbp!]
\caption{Decay widths of semilepton decay of double heavy baryons which induced by  $b\to c$.}\label{tab:ccq_semi}\begin{tabular}{|c|c|c|c|c|c|c|c}\hline\hline
channel & decay width($10^{-14}{\rm {GeV}}$)\\\hline
$\Gamma(\Xi_{bb}\to\Xi_{bc}) $ & $ 1.955\pm 0.685(T_{1}^{2},T_{2}^{2})\pm 1.673(s_{1}^{0},s_{2}^{0})$\\\hline
$\Gamma_L(\Xi_{bb}\to\Xi_{bc})$ & $1.728\pm 0.658(T_{1}^{2},T_{2}^{2})\pm 1.363(s_{1}^{0},s_{2}^{0})$\\\hline
$\Gamma_T(\Xi_{bb}\to\Xi_{bc}) $ & $ 0.227\pm 0.175(T_{1}^{2},T_{2}^{2})\pm 0.342(s_{1}^{0},s_{2}^{0})$\\\hline\hline
$\Gamma(\Omega_{bb}\to\Omega_{bc}) $ & $ 3.005\pm 0.780(T_{1}^{2},T_{2}^{2})\pm 4.932(s_{1}^{0},s_{2}^{0})$\\\hline
$\Gamma_L(\Omega_{bb}\to\Omega_{bc}) $ & $1.854\pm 0.670(T_{1}^{2},T_{2}^{2})\pm 2.363(s_{1}^{0},s_{2}^{0})$\\\hline
$\Gamma_T(\Omega_{bb}\to\Omega_{bc}) $ & $ 1.151\pm 0.382(T_{1}^{2},T_{2}^{2})\pm 2.562(s_{1}^{0},s_{2}^{0})$\\\hline\hline
$\Gamma(\Xi_{bc}\to\Xi_{cc}) $ & $4.174\pm 0.796(T_{1}^{2},T_{2}^{2})\pm 4.933(s_{1}^{0},s_{2}^{0})$\\\hline
$\Gamma_L(\Xi_{bc}\to\Xi_{cc}) $ & $ 3.260\pm 0.730(T_{1}^{2},T_{2}^{2})\pm 4.272(s_{1}^{0},s_{2}^{0})$\\\hline
$\Gamma_T(\Xi_{bc}\to\Xi_{cc}) $ & $ 0.914\pm 0.279(T_{1}^{2},T_{2}^{2})\pm 0.66(s_{1}^{0},s_{2}^{0})$\\\hline\hline
$\Gamma(\Omega_{bc}\to\Omega_{cc}) $ & $4.799\pm 1.095(T_{1}^{2},T_{2}^{2})\pm 4.385(s_{1}^{0},s_{2}^{0})$\\\hline
$\Gamma_L(\Omega_{bc}\to\Omega_{cc}) $ & $2.762\pm 0.676(T_{1}^{2},T_{2}^{2})\pm 2.931(s_{1}^{0},s_{2}^{0})$\\\hline
$\Gamma_T(\Omega_{bc}\to\Omega_{cc}) $ & $2.037\pm 0.867(T_{1}^{2},T_{2}^{2})\pm 1.454 (s_{1}^{0},s_{2}^{0})$\\\hline
\hline
\end{tabular}
  \end{table}

\begin{figure}[htbp!]
   \begin{minipage}[t]{0.45\linewidth}
  \centering
  \includegraphics[width=1.0\columnwidth]{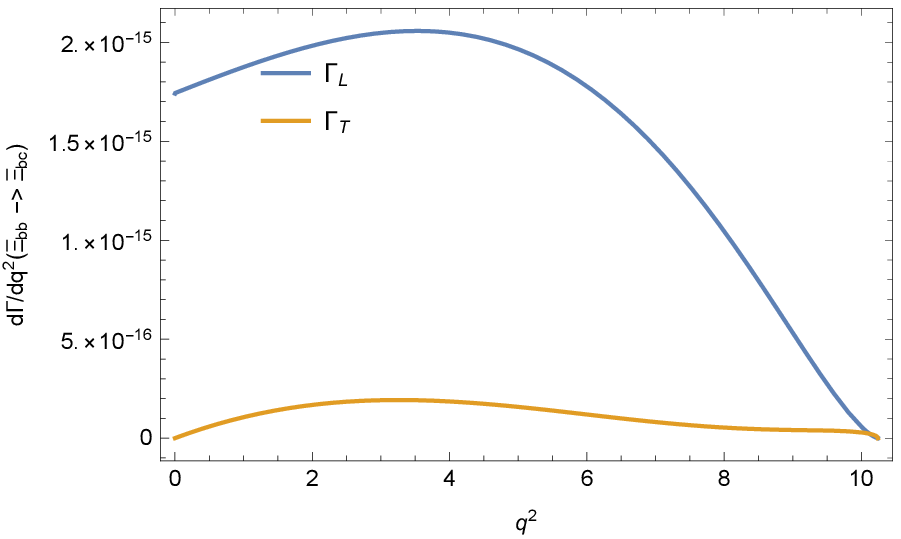}
    \end{minipage}
    \begin{minipage}[t]{0.45\linewidth}
  \centering
  \includegraphics[width=1.0\columnwidth]{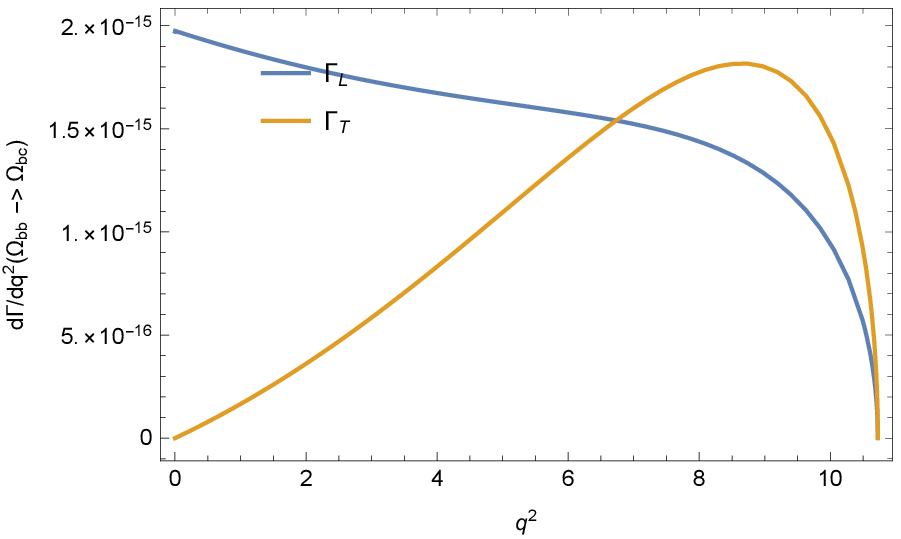}
    \end{minipage}
   \begin{minipage}[t]{0.45\linewidth}
  \centering
  \includegraphics[width=1.0\columnwidth]{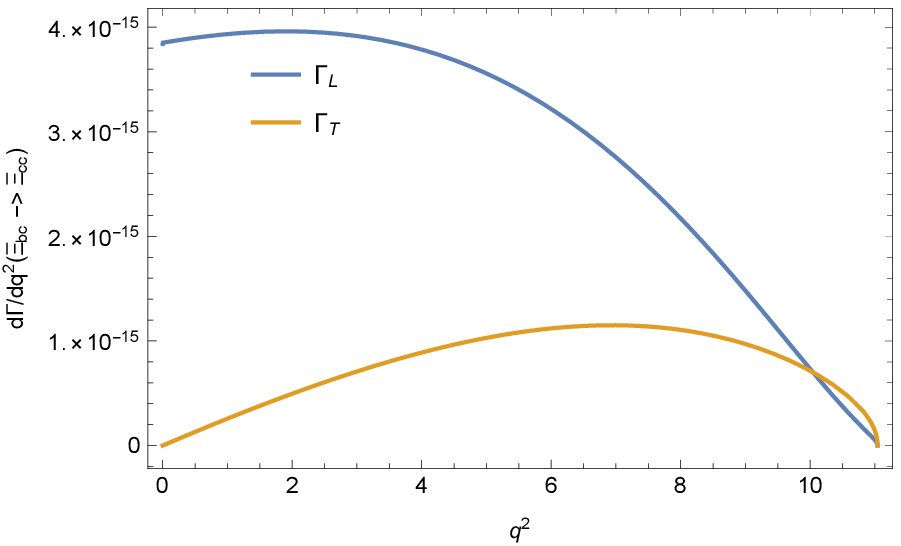}
    \end{minipage}
     \begin{minipage}[t]{0.45\linewidth}
  \centering
  \includegraphics[width=1.0\columnwidth]{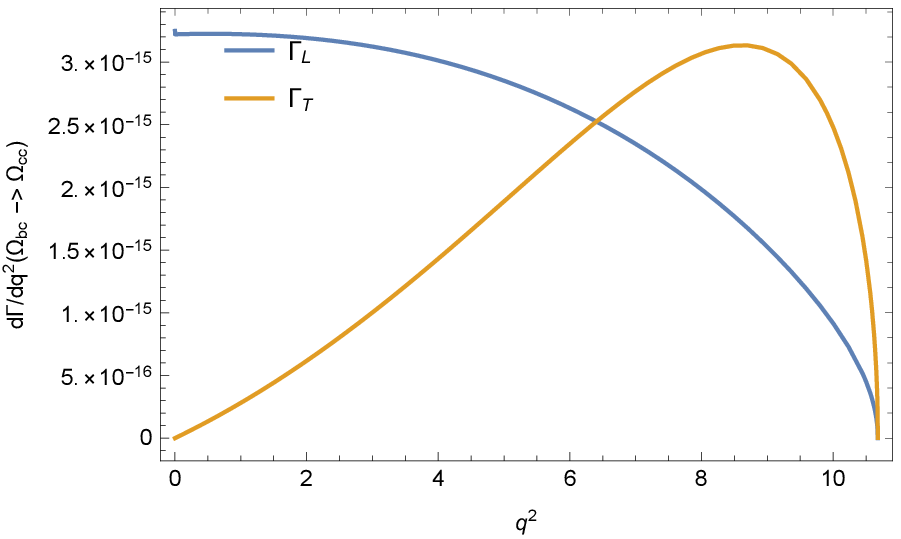}
    \end{minipage}
\caption{Differential decay widths of $\Xi_{bb}\to\Xi_{bc}$ (top left), $\Omega_{bb}\to\Omega_{bc}$
(top right), $\Xi_{bc}\to\Xi_{cc}$ (bottom left), and $\Omega_{bc}\to\Omega_{cc}$
(bottom right).}
\label{fig:dGamma}
\end{figure}

The differential decay widths are plotted in Fig.~\ref{fig:dGamma}.

\section{Summary}

In this work, we have investigated the $b\to c$ decay form factors
of doubly heavy baryons in QCD sum rules. For completeness, we have
also performed the analysis of pole residues, and as by-products,
the masses of doubly heavy baryons. Our predictions for the masses
are in good agreement with those of Lattice QCD and experimental data.
On the OPE side, contributions from up to dimension-5 and dimension-6
operators are respectively considered for the two-point and three-point
correlation functions. We have also considered the leading logarithmic
corrections for the Wilson coefficients of OPE, and it turns out that
these corrections are small. The obtained form factors are then used
to predict the corresponding semi-leptonic decay widths, which are
considered to be helpful to search for other doubly heavy baryons
at the LHC.

\section*{Acknowledgements}

The authors would like to
thank Prof. Wei Wang for constant help and encouragement. Z.-X. Zhao
is supported in part by scientific research start-up fund for
Junma program of Inner Mongolia University, scientific research start-up
fund for talent introduction in Inner Mongolia Autonomous Region,
and National Natural Science Foundation of China under Grant No. 12065020.


\end{document}